\documentclass[12pt,preprint]{aastex}

\shorttitle{QU Car}
\shortauthors{Linnell et al.}

\begin{document}

\title{Synthetic Spectrum Constraints on a Model of the Cataclysmic Variable QU Carinae\footnotemark[1]}
\footnotetext[1]
{Based on observations made with the NASA/ESA Hubble Space Telescope, obtained at the
Space Telescope Science Institute, which is operated by the Association of Universities 
for Research in Astronomy, Inc. under NASA contract NAS5-26555, and the NASA-CNES-CSA
{\it Far Ultraviolet Explorer}, which is operated for NASA by the Johns Hopkins University
under NASA contract NAS5-32985} 


\author{Albert P. Linnell$^2$, Patrick Godon$^3$, Ivan Hubeny$^4$, Edward M. Sion$^5$,
Paula Szkody$^6$, and Paul E. Barrett$^7$}

\affil{$^2$Department of Astronomy, University of Washington, Box 351580, Seattle,
WA 98195-1580\\
$^3$Department of Astronomy and Astrophysics, Villanova University,
Villanova, PA 19085\\
visiting at the Space Telescope Institute, Baltimore, MD.\\
$^4$Steward Observatory and Department of Astronomy,
University of Arizona, Tucson, AZ 85721\\
$^5$Department of Astronomy and Astrophysics, Villanova University,
Villanova, PA 19085\\
$^6$Department of Astronomy, University of Washington, Box 351580, Seattle,
WA 98195-1580\\
$^7$United States Naval Observatory, Washington, DC 20392\\
}

\email{$^2$linnell@astro.washington.edu\\
$^3$godon@stsci.edu\\
$^4$hubeny@as.arizona.edu\\
$^5$edward.sion@villanova.edu\\
$^6$szkody@astro.washington.edu\\
$^7$barrett.paul@usno.navy.mil\\
}

\begin{abstract}
Neither standard model SEDs nor truncated standard model SEDs fit observed spectra 
of QU Carinae with acceptable accuracy over the
range 900\AA~to 3000\AA. Non-standard model SEDs fit the observation set accurately.
The non-standard accretion disk models have a hot region extending from the white dwarf 
to $R=1.36R_{\rm wd}$,
a narrow intermediate temperature annulus, and an isothermal remainder to the tidal
cutoff boundary. The models include a range of $\dot{M}$ values
between $1.0{\times}10^{-7}M_{\odot}~{\rm yr}^{-1}$ and $1.0{\times}10^{-6}M_{\odot}~{\rm yr}^{-1}$
and limiting values of $M_{\rm wd}$ between $0.6M_{\odot}$ and $1.2M_{\odot}$.
A solution with $M_{\rm wd}=1.2M_{\odot}$ is consistent with an empirical mass-period relation. 
The set of models agree on a limited range of 
possible isothermal region $T_{\rm eff}$ values between 14,000K and 18,000K. The
model-to-model residuals are so similar that it is not possible to choose a best model.
The Hipparcos distance, 610 pc, is representative of the model results.
The orbital inclination is between $40\arcdeg$ and $60\arcdeg$. 

\end{abstract}




\section{Introduction}

Cataclysmic variables (CVs) are semi-detached binary stars in which a late-type main sequence star 
loses mass onto a white dwarf (WD) by Roche lobe overflow \citep{w1995}.
A sub-class, the NL systems (discussed below) are of interest because they are photometrically stable. 
It is believed that an analytic expression describes the radial temperature profile on the accretion
disk, leading to a corresponding spectral energy distribution (SED). Tests using stellar SEDs 
\citep{wade1988} produced unsatisfactory
results, but a theoretical model tailored to accretion disks \citep{hubeny1990} is available and might
be expected to lead to satisfactory results.
In this paper we show that synthetic SEDs, built on the
theoretically-prescribed model, do not fit the observed data, but non-standard empirical models fit the observations
closely.
 
In non-magnetic systems the mass
transfer stream produces an accretion disk with mass transport inward and angular momentum 
transport outward by viscous processes. The accretion disk may extend inward to the WD; the
outer boundary extends to a tidal cutoff limit imposed by the secondary star in the steady state case.
If the mass transfer rate is below a certain limit, the accretion disk is unstable
and undergoes brightness cycles (outbursts), and if above the limit, the accretion disk is stable against 
outbursts \citep{s1986,c1993,w1995,o1996}.
The latter case objects are called nova-like (NL) systems; their accretion disks are nearly fully ionized
to their outer (tidal cutoff) boundary and this condition suppresses dwarf nova outbursts.
NL systems are of special interest because they are expected to have an accretion disk radial temperature
profile	given by an analytic expression \citep[eq.5.41]{fr92}(hereafterFKR) thereby defining
the so-called standard model. 
 
Based on reports of QU Carinae as an emission-line object \citep{st1968}, \citet{gil1982} (hereafter GP)
undertook a spectroscopic study to determine orbital characteristics of the system. They determined an
orbital period of 10.9 hours. No secondary star spectrum was detected, indicating that the spectrum is
dominated by light from an accretion disk or the primary star. Balmer line emission is relatively weak.
This evidence in turn suggested a high
rate of mass transfer and a correspondingly bright system. It also suggests that QU Car is a NL-type system.
We discuss known parameters of QU Car in more detail in \S4.

High speed photometry confirmed erratic variations of 0.1-0.2 mag. on time scales of minutes, as reported by
\citet{schild1969}. \citet{kn1994} used $IUE$ spectra to study time variability of the UV resonance lines.
They found that there are short-time-scale variations with a possible correlation with orbital period, 
but the data are 
insufficient to confirm the orbital nature of the variations.

\citet{har2002}	obtained $Hubble~Space~Telescope$ $STIS$ spectra of QU Car at three epochs. Blueshifted absorption
in N V and C IV provides evidence of an outflow with a maximum velocity of $\approx 2000$ km~${\rm s}^{-1}$.
Further analysis of these spectra is in \citet{drew2003}. The latter authors suggest that the QU Car secondary 
is a carbon star, and that QU Car is the most luminous NL system known, possibly at a distance of
${\approx}2000$ pc.

Our objective in this study is to use synthetic spectra, based on a physical model of QU Carinae, to constrain
properties of this interesting system.

\section{Interstellar reddening and the Hipparcos distance}

\citet{ver1987} determined $E(B-V)=0.1{\pm}0.03$ for QU Car from the 2200\AA~interstellar medium(ISM) 
feature. \citet{drew2003} used the $STIS$ spectra (the same applied in this investigation) to determine
a neutral hydrogen column density of $(6{\pm}1){\times}10^{20}~{\rm cm}^{-2}$. For a standard gas-to-dust 
ratio of
$N({\rm H1)}/E_{B-V}=5.8{\times}10^{21}~{\rm mag}^{-1}~{\rm cm}^{-2}$ \citep{boh1978}, the corresponding
reddening is $E(B-V)=0.1{\pm}0.015$, in agreement with \citet{ver1987}. 
In \S8, we model the ISM atomic and molecular hydrogen absorption
lines and find a hydrogen column density of
$N({\rm H1})=2.5 \times 10^{20}$cm$^{-2}$, slightly lower than the value
obtained by Drew et al., but within the range of acceptable values
for a reddening of E(B-V)=0.1 (the Bohlin et al. empirical law has
a very large scatter).
We adopt $E(B-V)=0.1$.

The Hipparcos parallax \citep{perry1997} of $1.64{\pm}1.50(\rm mas)$, corresponds to a distance of 610 pc, 
with $1{\sigma}$ errors giving a range between 318 pc and 7143 pc. Thus the Hipparcos parallax does not 
impose a
strong constraint.

\section{The $FUSE$, $STIS$, and $IUE$ spectra}

The $FUSE$ and $STIS$ spectra were obtained at a time separation of about three years.
A conversion factor was anticipated to match the two spectra, and
we found it necessary to multiply the $STIS$ data by the factor 1.263 to match the
flux level of the $FUSE$ spectrum in the overlap region. Note also that the $STIS$ spectrum used in
our analysis is a combination of three exposures obtained over an interval of
nearly one year.
The data were dereddened to E(B-V)=0.1 (\S2) after matching 
the flux levels. 

The $FUSE$ spectrum of QU Car was obtained 
through the 30"x30" LWRS Large Square Aperture, in TIME TAG mode
and was obtained on 23 April 2003, dataset D1560102000, PI Name: L. Hartley. 
The data were processed with CalFUSE version 3.0.7 \citep{dixon2007} which automatically handles 
event bursts. Event bursts are short periods during an exposure when high count rates are registered 
on one or more detectors. The bursts exhibit a complex pattern on the detector; their cause is as 
yet unknown (it has been confirmed that they are not detector effects). 
The main change from previous versions of CalFUSE is that now the data
are maintained as a photon list (the intermediate data file - IDF) throughout
the pipeline. Bad photons are flagged but not discarded, so the user can examine
the filter and combine data without re-running the pipeline.
 
The spectrum actually consists of 3 exposures, each covering a portion of a single
orbit.   
These exposures, totaling 10,748 sec of good exposure time, were combined (using the IDF files).  
Consequently,  
the $FUSE$ spectrum has a duration of about 0.27 of the period of QU Car.
The spectral regions covered by the $FUSE$ 
spectral channels overlap, and these overlap regions are then used to
renormalize the spectra in the SiC1, LiF2, and SiC2 channels to the flux in
the LiF1 channel. We combined the individual channels to create a
time-averaged spectrum with a linear $0.05$\AA\, dispersion, weighting
the flux in each output datum by the exposure time and sensitivity of the
input exposure and channel of origin. 
In this manner we produced a final spectrum that covers the
full {\it{FUSE}} wavelength range $905-1187$\AA. 
  
The $STIS$ spectra used in this study are those obtained in 2000 by \citet{har2002}.
The three $STIS$ spectra were processed with CALSTIS version 2.22. 
The $HST/STIS$ spectra of QU Car were obtained in TIME TAG 
operation mode using the FUV MAMA detector.  
The observations were made through the 0.2x0.2 ${\rm arcsec}^2$ aperture 
with the E140M optical element. Additional details are in \citet{har2002}.
The $STIS$ spectra have a total
duration of 7500 sec or about 0.2 of the period of QU Car.
The $STIS$ spectrum, centered on the wavelength 1425\AA, 
consists of 42 echelle spectra that we extracted and combined
to form a single spectrum extending from about 1160\AA~to 1710\AA~and with a
resolution of 0.1\AA.  

$IUE$ spectra extend the wavelength coverage
to 3000\AA~from the $STIS$ limit of 1710\AA, thereby providing a significantly enhanced constraint
on a synthetic spectrum model. 
The $IUE$ archive contains 6 SWP spectra and two LWP spectra of QU Car for the dates
26-27 June 1991, among others. Because of the intrinsic variability of the system, we have restricted
our $IUE$ data to that time interval.
 
We chose a single pair of $IUE$ spectra, SWP41927 and LWP20700,
which were taken in immediate sequence on 27 June 1991. 
The close temporal sequence minimizes the possibility of source changes affecting the observations.
We dereddened the spectra individually 
and merged them
with the IUEMERGE facility, with the `concatenate' option. 
We used an expanded ordinate scale to study the overlap of the $STIS$ and merged $IUE$ spectra
and found that visual inspection provided a sensitive test of accurate superposition.
In this way we found that the optimum divisor to superpose the 
merged spectrum on the scaled and dereddened $STIS$ spectrum was 0.85. 

Figure~1 shows the $FUSE$ and the $STIS$
spectra of QU Car before dereddening. Note the very large number of interstellar lines in 
the $FUSE$ spectrum. 
In Table~1 we identify the prominent absorption lines of the $FUSE$
spectrum. 
The wavelength tabulations (column 2) are the theoretical wavelengths.
For clarity we did not include the multitude of ISM lines in the table.
A complete list of ISM lines (hydrogen + metals) can be found
e.g. in \citet{sem1999} and \citet{sem2001}.
Table~2 identifies the absorption lines in the $STIS$ spectrum.
The ISM lines of the $STIS$ spectrum have been analyzed in
Drew et al. (2003), and the broad absorption lines
(C\,{\sc iii}$\lambda$1176, N\,{\sc v}$\lambda$1240,
O\,{\sc v}$\lambda$1371, Si\,{\sc iv}$\lambda$1398,
C\,{\sc iv}$\lambda$1549, He\,{\sc ii}$\lambda$1640)
have been studied in detail for each $STIS$ exposure
separately in Hartley et al. (2002). The broad lines
are seen in emission with a velocity broadening of several
thousand km/s; an absorption component is superposed 
with a velocity broadening of a few hundred km/s.

\section{Parameters adopted initially}

The orbital period of 10.9 hours was determined by GP, using 
radial velocities
from the ${\lambda}{\lambda}4630-4660$ emission feature and the ${\lambda}4686$ HeII emission
line. There is no detectable orbital light modulation \citep[][GP]{schild1969}, suggesting
an inclination $i{\le}60{\arcdeg}$. The presence of doubled emission lines (GP)
suggests $i{\ge}30{\arcdeg}$. 

Based on the main sequence mass-radius relation of \citet{lacy1977} and the usual period-mean density
relation \citep{rob1976}, GP obtained an approximate $M_2 (=M_{\rm s})=1.3M_{\odot}$.
Note that this determination does not depend on an explicit knowledge of $i$.
GP spectrophotometry sets a limit of $\approx1\%$ on the depth of any absorption feature 
from the secondary star in the QU Car
optical spectrum. Comparison with the $\approx10\%$ depth of absorption features in G-K standards
(appropriate to the secondary star) establishes that the disk and primary of QU Car must be
radiating at $\gtrsim10L_\odot$. The strong He II $\lambda4686$ emission, weaker He II $\lambda4542$
emission, and the absence of He I lines at $\lambda4387$, $\lambda4471$, and $\lambda4713$ 
suggests a temperature in the
line-emitting region in excess of 20,000K. This is consistent with a fully ionized accretion disk
expected for a NL system.
GP determine a minimum distance to QU Car of 500 pc, based on $m_B=11.2$ for
the system and standard assumptions for interstellar absorption. From the $M_V-P_{\rm orb}$ relation
of \citet[Figure 4.16]{w1995}, with $M_V=2.4$ and interstellar absorption of 0.3 mag. (\S2), we
obtain a minimum distance that agrees with the GP value. 

A coarse estimate of the mass transfer
rate is possible using the study by \citet{s1989}. From Smak's Figure~2 we obtain 
$\dot{M}\sim5\times10^{-7}M_{\odot}{\rm yr}^{-1}$. 
Later, \citet{s1994} slightly revised the calculated 
mass transfer rates. This value is roughly consistent with the mass transfer rate-orbital period
relation of \citet[Figure 21]{pat1984}.
\citet{drew2003}
find that the QU Car C/He abundance may be as high as 0.06, an order of magnitude higher than
the solar ratio, and the C/O abundance is estimated to
be greater than 1. 
Based on a power law fit to merged $IUE$ spectra, and a possible cutoff of the spectrum at 228\AA~(the HeII series limit),
\citet{drew2003} suggest a mass transfer rate of 
$\dot{M}=8.0{\times}10^{-8}~{M}_{\odot}{\rm yr}^{-1}$ if QU Car is at a distance of 500 pc and
$\dot{M}=1.0{\times}10^{-6}~{M}_{\odot}{\rm yr}^{-1}$ if the distance is 2000 pc.

GP cite the
uncertainty in the $M_{\rm s}$ value and the lack of knowledge of $i$, and so do not derive a
value of $M_{\rm wd}$. However, some constraint on $M_{\rm wd}$ is possible as shown by the following
discussion.

From the standard expressions for radial velocity amplitude \citep[p.360, eq. 60]{sm47}, we obtain
\begin {equation}
1+\frac{1}{q}=\frac{2{\pi}{\rm sin}i}{K_{\rm wd}P}D,
\end {equation}
where we assume zero orbital eccentricity; the mass ratio $q=M_{\rm s}/M_{\rm wd}$, $i$ is the orbital
inclination, $K_{\rm wd}$ is the WD radial velocity amplitude, $P$ is the orbital period, and $D$
is the component separation. From Kepler's third law,
\begin {equation}
1+\frac{1}{q}=\frac{4{\pi}^2}{GP^2}\frac{1}{M_{\rm s}}D^3.
\end {equation}
From these two equations we have
\begin {equation}
D^2=GM_{\rm s}\frac{P}{2{\pi}}\frac{{\rm sin}i}{K_{\rm wd}}.
\end {equation}
GP determine a  value $K_{\rm wd}=115~{\rm km}~{\rm sec}^{-1}$ from He II emission lines (insecure because
emission lines may not trace the WD motion accurately). GP derive
an orbital period of $0.4542^d$. With these parameters determined by observation, an adopted value of $i$
and an assigned value of $M_{\rm s}$ determines $D$ from equation (3), $q$ from equation (1), and 
$M_{\rm wd}$ from $M_{\rm wd}=M_{\rm s}/q$. Thus, a given $i$ has an associated contour in the mass-mass plane
that determines $M_{\rm wd}$ values corresponding to assigned $M_{\rm s}$ values.

Figure~2 illustrates the principle. The WD axis is bounded above by the Chandrasekhar mass. 
The diagonal line represents the case of equal 
WD and secondary
star masses; systems represented by points above this line are excluded by dynamical instability. 
(Mass transfer from a more-massive secondary causes the period to shorten, the separation of components 
and the secondary component Roche lobe to shrink, and so accelerates the mass transfer.)
The value of $i$ cannot be greater than about $60\arcdeg$ to avoid eclipses. 
No models with $i$ as small as $30\arcdeg$ are permitted (for $q=1.0$ the masses of both stellar components
then are larger than the Chandrasekhar mass and for $q<1.0$ the WD mass becomes even larger).
A $M_{\rm s}>1.30M_{\odot}$ is not permitted because the secondary must lie on or above the ZAMS
mass-radius relation \citep{lacy1977}.
The empirical mass-period relation of \citet[eq. 2.100]{w1995} gives $M_{\rm s}=1.29M_{\odot}$.  
A value $M_{\rm s}=1.20M_{\odot}$, with $M_{\rm wd}=1.20M_{\odot}$ is permitted with $i=38\arcdeg$.
The uncertainty in $K_{\rm wd}$ leads us to adopt $i=40\arcdeg$, $M_{\rm wd}=M_{\rm s}=1.20M_{\odot}$ as an 
appropriate case for simulation.

\section{Calculation of system models}

Our calculation of system models uses the BINSYN suite \citep{linnell1996,linnell2007a}.
The word 'model', as used in this section, has two context-dependent meanings. In one context it refers
to a listing of the run of physical parameters in an annulus, or to properties of the collection of annuli comprising
an entire accretion disk. In a different context it refers to the array of output synthetic spectra from BINSYN. 
In the latter context, 
orbital phase-dependent synthetic spectra  
separately of the system, the WD, 
the secondary star, the 
accretion disk, and the
accretion disk rim, all produced in a single run, constitute a BINSYN model of a CV system. 
The synthetic spectra include effects of eclipse
and irradiation.
 
A BINSYN accretion disk model uses an array of
annulus models calculated with the program TLUSTY \citep{hubeny1988,hubeny1990,hl1995,hh1998} and
corresponding synthetic spectra calculated with program SYNSPEC	\citep{hlj1994}. 
Table~3 is an illustration of an accretion disk model for $M_{\rm wd}=1.2M_{\odot}$ and
$\dot{M}=6.0{\times}10^{-7}~{M}_{\odot}{\rm yr}^{-1}$.  
The 26 TLUSTY 
annulus models cover the interval from the
WD equator to the tidal cutoff boundary. The cutoff boundary is calculated from \citet[p. 57, eq. 2.61]{w1995}. 
Note that the annulus radii in column~1 are measured in units of 
a zero temperature WD. Successive columns provide some of the data available in the TLUSTY output for a
given annulus model; each row 
describes
a separate annulus model. 
The annuli all are H-He models. (i.e., these are the only explicit atoms; 
the next 28
elements are treated implicitly. See the TLUSTY manual for details.~\footnote{http://nova.astro.umd.edu}) 
Convection was ignored.
The models converged for all of the Table~3 annuli. In some instances, for other $\dot{M}$ values,
the outermost annuli failed to converge and we used so-called gray models (see the TLUSTY manual).

The $m_0$ column lists the mass per unit area (${\rm gm}~{\rm cm}^{-2}$)
between the central plane and the upper boundary, Calculation of an annulus model specifies an initial 
uppermost layer
in ${\rm gm}~{\rm cm}^{-2}$. 
Our calculation used $10^{-4}~{\rm gm}~{\rm cm}^{-2}$ for the initial
layer.
An annulus model typically may have 70 depth
points. 
$T_c$ is the central plane temperature, log~$g$ is the value of that parameter at 
optical depth near 0.7,
$z_H$ is the height, in cm, of the optical depth 0.7 level above the central plane, and ${\tau}_{\rm Ross}$ is the
Rosseland optical depth at the central plane. 

SYNSPEC is used to calculate a synthetic spectrum
for each TLUSTY annulus, with specified spectral resolution. 
The line list includes data for the first 30 periodic table elements.
Next, BINSYN is used with its specified number of
annuli (different from the set of TLUSTY annuli) to calculate the various output synthetic spectra by integration over
the array of BINSYN annuli for the accretion disk and by corresponding integrations over the stellar photospheric
grids for the latter objects. 
Synthetic spectra for the individual BINSYN annuli follow
by interpolation among the array of SYNSPEC spectra of the TLUSTY annuli.
It is at this level that information is incorporated on orbital inclination, eclipse effects, radial velocity 
effects, etc.,
as needed for individual photospheric segments of all objects in the system.

Calculation of a synthetic spectrum corresponding to a given $\dot{M}$ model is a 
computationally-intensive process and compromises 
are necessary in deciding
how many TLUSTY annulus models to calculate to achieve acceptable accuracy in the final system synthetic spectrum. 
We believe our representational
accuracy is high, but we cannot be sure that the fit in, e.g., Figure~8 (following) could not be
improved with a finer grid of annulus synthetic spectra for the TLUSTY annulus models and a larger number of 
BINSYN annuli.
We note that this entire process must be repeated for each value of $\dot{M}$ and in
turn for each value of $M_{\rm wd}$.
 
In general, the WD synthetic spectrum contribution to the system synthetic spectrum is a strong function of
the adopted WD $T_{\rm eff}$. Since the WD radius varies with the adopted WD $T_{\rm eff}$, it makes a
difference whether we use the zero temperature radius or the corrected radius in calculating annulus
radii, in BINSYN, measured in units of the WD radius.
We must use the corrected radius with BINSYN, designated
by the symbol $r_{\rm wd}$, since the output spectra must describe the WD contribution for its $T_{\rm eff}$
actually adopted.
 
The WD $T_{\rm eff}$ likely will be unknown in advance of the model calculation.
We determine what annulus models to calculate with TLUSTY by using a fixed set of multiples of
the zero temperature WD radius; we stress that, if we were to use the fixed multiples procedure with a 
variable WD radius, 
it would be
nearly prohibitive to recalculate the entire array of TLUSTY annuli with each new approximation to
an adopted WD $T_{\rm eff}$. Such iterative recalculation is unnecessary; BINSYN temperature-wise 
interpolation among
the fixed (with $r_{\rm wd}=r_{\rm wd,0}$) TLUSTY annulus models leaves the accretion disk 
synthetic spectrum essentially 
unaffected as the WD $T_{\rm eff}$-dependent radius
changes. 
We adopt solar composition for the annulus models.
In the present application the WD contribution is too small to matter,
but it is important to retain the distinction, as this paper does.

Our initial model, at all mass transfer rates,
adopted a standard model \citep[FKR,][]{linnell2007a}. 
A standard model assumes
the accretion disk is in hydrostatic equilibrium, that the accretion disk material is very nearly
in Keplerian rotation, and that a single viscosity parameter applies to the entire accretion disk.
All of our TLUSTY annulus models adopted a viscosity parameter \citep{ss1973} $\alpha=0.1$.

\section{QU Carinae models with $M_{\rm wd}=1.20M_{\odot}$, $q=1.00$}

The long orbital period and the weak Balmer line emission, suggesting a NL-type system (GP), in turn
indicate a hot WD. The radius of a $1.2M_{\odot}$ WD is slightly sensitive to the photospheric
temperature. 
Table 12C-H-S.dat of \citet{panei2000}, for homogeneous Hamada-Salpeter models, lists a radius 
$r_{\rm wd,0}=5.6\times10^{-3}R_{\odot}$ for a zero temperature, carbon, $1.2M_{\odot}$ WD. 
Based on an adopted 55,000K
WD, we used Figure~4a of \citet{panei2000} to determine a corrected radius of 
$6.7{\times}10^{-3}R_{\odot}$.
We will find that the WD makes an insignificant contribution to the system luminosity; the only
impact the WD $T_{\rm eff}$ has is indirect--temperatures in the accretion disk at specified
multiples of the WD radius vary with the $T_{\rm eff}$-dependent
WD radius.

A smaller assumed WD $T_{\rm eff}$ produces a smaller $R_{\rm wd}$ and a correspondingly larger accretion
disk $T_{\rm eff}$ in the innermost rings (FKR) since they then are in a deeper potential well. However, 
this effect is smaller than the temperature effect, on all annuli, of different
assumed mass transfer rates. For uniformity in comparing different assumed mass transfer rates we
arbitrarily adopt a fixed WD $T_{\rm eff}$=55,000K.
The tidal cutoff radius of the accretion disk is $150R_{\rm wd}$ for the adopted 
WD $T_{\rm eff}$, and also equals $0.998R_{\odot}$.
(Contrast with the same tidal cutoff radius, measured in units of the zero temperature WD radius,
in the last line of Table~3).
The BINSYN model uses 45 division circles on the accretion disk, with the first division at the WD 
equator, the next two at $1.1806R_{\rm WD}$, and $1.3611R_{\rm wd}$,
and the last at the tidal cutoff boundary. 
 
We tested models with
$\dot{M}=0.1,0.3,0.6,1,3,6,~{\rm and}~10{\times}10^{-7}~{M}_{\odot}{\rm yr}^{-1}$.
In the first case a standard model produces an accretion disk temperature profile whose outer
annuli are cooler than 6000K. This model is unstable \citep{smak1982} and must be rejected.
In all remaining cases, the standard model produces an accretion dsik profile with a spectral
gradient that is too large.

Figure~3 illustrates the problem for $\dot{M}=6.0{\times}10^{-7}~{M}_{\odot}{\rm yr}^{-1}$; 
other values of $\dot{M}$ produce comparable plots after applying appropriate normalizing factors.
In all cases in this paper, normalizing factors have been determined by eye estimates.
In Figure~3 the peak flux of the WD, near 1000\AA, is less than 0.01 of the system flux; the WD contribution
to the system spectrum is negligible. 
Note that the synthetic spectrum consists of a continuum spectrum plus absorption lines of
H and He II. (He II but not He I because He II is a single electron ion.) 
The spectral gradient problem persists if $i$ is arbitrarily changed to 50 deg. 
For reasons discussed 
in \S10, the
$\dot{M}=1.0{\times}10^{-6}~{M}_{\odot}{\rm yr}^{-1}$ rate is a maximum, so our range of $\dot{M}$
values cover the full range of possible choices for this $M_{\rm wd}$.

We conclude that, for this WD and secondary star mass, no standard model SED provides an
acceptable fit to the observational data. We believe the spacing between our $\dot{M}$ values is small
enough to extrapolate our conclusion to any $\dot{M}$ value not actually calculated.

Truncated standard models are of interest because the inner accretion disk can be elided by radiation from
a hot WD or by interaction with a magnetic field associated with the WD. We ignore whether
either process is likely for QU Car and restrict consideration to whether a truncated
accretion disk can provide a SED that fits the observations with acceptable accuracy.

A truncated accretion disk model, with the residual disk
on the standard model temperature profile, should have a lower spectral
gradient and so is a possible alternative.
We have calculated truncated models for the last six mass transfer rates listed above (The
model with the lowest rate of mass transfer failed because of its temperature behavior at large radii). 
Truncation cuts off the annuli that contribute most to the far UV.
In the six largest $\dot{M}$ cases, the best truncation compromise either still had a too large 
spectral gradient or failed
to represent the $FUSE$ spectral region with acceptable fidelity. 

Figure~4 illustrates the fit
for $\dot{M}=6.0{\times}10^{-7}~{M}_{\odot}{\rm yr}^{-1}$, and a truncation radius of 
$R_{\rm trunc}=22.5R_{\rm wd}$. The overall fit is an improvement over Figure~3, but the fit
to the $FUSE$ spectrum is unsatisfactory and the spectral gradient still is too large. A
larger truncation radius would improve the spectral gradient problem but make the fit to the
$FUSE$ spectrum worse. A smaller truncation radius has a still more discrepant spectral gradient.
Other values of $\dot{M}$ are unsatisfactory for similar reasons.

Having demonstrated that no standard model SED or truncated standard model SED fits the observational
data, we consider non-standard model SEDs.
The too-large spectral gradient can be reduced with a cooler (average) accretion disk. However,
the flux maximum in the $FUSE$ spectrum near 1000\AA~ requires the presence of a hot region
on the accretion disk. This suggests reducing the temperature of the disk model at intermediate radii.

It is important to note that a modification of the temperature profile removes the direct connection
to a specific mass transfer rate. Not only is the empirical profile disconnected from the standard
model, the total flux from the empirical model differs from the flux of the parent standard model.
We conducted very extensive preliminary experiments and found that a steep drop in the accretion disk temperature,
followed by a much lower temperature gradient than in the standard model gave a fairly good fit to
the observed spectra. However, we determined that further improvement was possible.

It is desirable to preserve model connection to some mass transfer rate. On the standard model, 1/2
of the potential energy liberated by mass falling from the L1 point 
(located on the secondary star photosphere) to the surface of the WD
is converted to emitted radiation by the accretion disk (FKR). The remaining half either is liberated in a boundary
layer, goes into spinning up the WD, or is deposited elsewhere, as in a wind. Based on energy
considerations, a non-standard model accretion disk still can be associated with a mass transfer rate
if roughly 1/2 of the total liberated potential energy appears as flux from the accretion disk.

With that constraint in mind, we performed further tests. In the standard model the maximum accretion
disk temperature occurs at $R(T_{\rm max})=1.3611R_{\rm wd}$. The region closer to the WD is the boundary
layer region (discussed further in \S10). Standard model temperatures, for the $\dot{M}$ values 
we are considering, are so high
at the temperature maxima that only the Rayleigh-Jeans tail contributes to the observed spectra 
from those regions. (If, in addition, as described above, 1/2 of the total liberated potential 
energy were deposited in
the boundary layer region the $T_{\rm eff}$ there would become much higher.) 
The
bolometric flux contribution, per unit area, of the innermost annuli far exceeds that of more
remote parts of the accretion disk. Consequently we have preserved the standard model temperature maxima
of the boundary layer region and have modified the temperatures of the remaining annuli.
This choice preserves a connection with a particular $\dot{M}$ through the sensitivity of the total
accretion disk flux to the contribution of the highest temperature annuli. For convenience of
nomenclature, we designate the region between the WD and the accretion disk $R=1.3611R_{\rm wd}$
as the boundary layer even though this term has a slightly different connotation in the literature.

It has
proved possible to achieve excellent fits by adopting an isothermal profile over almost all of the
accretion disk while preserving a narrow region adjacent to the boundary layer, discussed in more 
detail below, and assigned an intermediate
temperature value. The narrow region is important in achieving a fit to the short wavelength part
of the $FUSE$ spectrum. The highest temperature boundary layer region, important in the total
accretion disk energy budget, contributes to the flux shortward of 1000\AA~in the observed spectrum 
by an amount that
varies slowly with its temperature (the contribution is in the Rayleigh-Jeans tail). Changing the
boundary layer $T_{\rm eff}$ from its standard model value produces appreciable changes in the
system flux (thereby disconnecting from the adopted $\dot{M}$ designation) without producing a
sensitive fit to the observed FUV flux. The narrow intermediate temperature region provides a
reasonably sensitive means to produce a good fit to the observed FUV flux. In all of our models,
the contribution of the intermediate temperature region near 1000\AA~also is in its Rayleigh-Jeans
tail.
 
The range of isothermal values that we have tested and that produce acceptable fits 
for this $M_{\odot}$, is limited. The range
extends from 15,000K at $\dot{M}=1.0{\times}10^{-7}~{M}_{\odot}{\rm yr}^{-1}$
to	17,000K at
$\dot{M}=1.0{\times}10^{-6}~{M}_{\odot}{\rm yr}^{-1}$ (higher isothermal temperatures produce a too-large
spectral gradient). 

Figure~5 presents a sample fit to the observed spectra, again for 
$\dot{M}=6.0{\times}10^{-7}~{M}_{\odot}{\rm yr}^{-1}$. The isothermal region has a temperature of 
17,000K.
In this comparison we have calculated a line
spectrum with a resolution of 0.02\AA. The resolution of the $FUSE$ spectrum is 0.05\AA. 
Our choice of synthetic spectrum resolution was to permit an accurate correction for ISM absorption,
described in \S8.
Figure~6
shows details of the fit to the $FUSE$ spectrum and the FUV end of the $STIS$ spectrum. Extensive
absorption lines of the ISM complicate the comparison. Further consideration of this topic is in \S8.

Figure~7 shows the fit to the $STIS$ spectrum. The failure to fit Ly~$\alpha$, as we show subsequently,
is rectified by including ISM absorption.
The 1200\AA~to 1500\AA~interval is affected by the ISM.
With the exception of not fitting the He II emission line, the fit in the 1560\AA~to 1710\AA~interval, shown
in Figure~8, generally is very good. 
Finally, Figure~9 shows
the accurate fit to the red end of the SWP spectrum and the blue end of the LWP spectrum. 

The bolometric luminosity for the $\dot{M}=6.0{\times}10^{-7}~{M}_{\odot}{\rm yr}^{-1}$ standard model
accretion disk is $1.01{\times}10^{37}~{\rm ergs}~{\rm sec}^{-1}$. The bolometric luminosity of the
empirical model is $3.95{\times}10^{36}~{\rm ergs}~{\rm sec}^{-1}$. 
Figure~5--Figure~9 are representative of corresponding plots for the four largest mass transfer rates for this 
WD and secondary star mass.	

The system parameters for this $M_{\rm wd}$ are in Table~4. The default BINSYN model (before imposing a non-standard
accretion disk temperature profile)
assigns standard model properties (FKR) at the various annulus radii, including the annulus semi-thickness. Those
properties vary slightly with $\dot{M}$ but the tabulated value of $H$, within the thickness resolution listed, 
is appropriate for the range of
$\dot{M}$ included in this study. Because our empirical models preserve a connection to corresponding standard
model $\dot{M}$ values, the $H$ value remains the same in the empirical models.

\section{QU Carinae models with $M_{\rm wd}=0.6M_{\odot}$, $q=0.83$}

We wish to test the possibility of acceptable standard model SED fits to the observed spectra over the
full range of possible $M_{\rm wd}$ and $\dot{M}$ values. 
Figure~2 indicates that the smallest possible $M_{\rm wd}$ for this system is about $0.6M_{\odot}$.
Again from Figure~2, $M_{\rm wd}=0.6M_{\odot}$
and $M_{\rm s}=0.5M_{\odot}$ fall on the $i=55\arcdeg$ contour.
From Figure 4a of \citet{panei2000}, for a $M_{\rm wd}=0.6M_{\odot}$ with
$T_{\rm eff}=55,000$K, we find $R_{\rm wd}=1.59{\times}10^{-2}R_{\odot}$. The corresponding radius of a zero
temperature WD is $1.21{\times}10^{-2}R_{\odot}$. For models with a range of $\dot{M}$ values, and the WD 
$T_{\rm eff}$=55,000K, the tidal cutoff radius of the
accretion disk is $53R_{\rm wd}$, which also is $0.840R_{\odot}$.

We tested models with 
$\dot{M}=0.3,0.6,1,3,6,~{\rm and}~10{\times}10^{-7}~{M}_{\odot}{\rm yr}^{-1}$.
The outer part of a standard model accretion disk for the first mass transfer rate is cooler than 6000K and
is unstable \citet{smak1982}. That model must be rejected.
In all remaining cases, the standard model produces an accretion disk profile with a spectral 
gradient that is
too large. The effect is similar to the $M_{\rm wd}=1.20M_{\odot}$ cases. 
Figure~10, illustrating the standard model
for $\dot{M}=3.0{\times}10^{-7}~{M}_{\odot}{\rm yr}^{-1}$, 
is representative
of the standard model synthetic spectrum fits. 
As with the $M=1.20M_{\odot}$ case, we find that no standard model
SED fits the observed spectra.

Figure~11 presents a truncated model for $\dot{M}=3.0{\times}10^{-7}~{M}_{\odot}{\rm yr}^{-1}$.
The truncation radius is $R_{\rm trunc}=9.91R_{\rm wd}$.
The overall fit is a big improvement over Figure~10 but the fit to the $FUSE$ spectrum is unsatisfactory.
The same type defect is present for the other $\dot{M}$ values. We conclude that no truncated model
provides a SED which fits the observed spectra satisfactorily. 
As with the $M_{\rm WD}=1.20M_{\odot}$ case,
it proves possible to produce good quality synthetic SED fits for the four largest mass transfer 
rates we tested. Figure~12 illustrates the result for $\dot{M}=3.0{\times}10^{-7}~{M}_{\odot}{\rm yr}^{-1}$.
Fits for other mass transfer rates are of similar quality.
Figure~13 presents the $FUSE$ fit, and Figure~14 the $STIS$ fit. Figure~15 shows the expanded scale fit,
1560\AA~to 1710\AA. 
Figure~16 shows the fit to the $IUE$ spectra and the accurate connection to the red end of the $STIS$
spectrum.
Parameters of the model for this $M_{\rm wd}$ are in Table~5. The comments describing Table~4 also apply
to Table~5. 

In summary, over the full range of permissible $M_{\rm wd}$ masses in this system, no acceptable $\dot{M}$
has an associated standard model SED that fits the observed spectra with acceptable accuracy.
The same conclusion follows for truncated standard models. As with the $M_{\rm wd}=1.2M_{\odot}$
case, all permissible $\dot{M}$ cases have an associated empirical accretion disk model that
provides an accurate fit to the set of observed spectra.
Further discussion is in \S10

\section{Inclusion of a model of the ISM}

We have developed software to simulate the ISM and have used this routine to determine
ISM properties in the line of sight to QU Car. 
A custom spectral fitting package is used to estimate the temperature and
density of the interstellar absorption lines of atomic and
molecular hydrogen.  The ISM model assumes that the temperature, bulk
velocity, and turbulent velocity of the medium are the same for all atomic
and molecular species, whereas the densities of atomic and molecular
hydrogen, and the ratios of deuterium to hydrogen and metals (including
helium) to hydrogen are adjustable parameters. The model uses atomic data of
\citet{morton2000, morton2003}
and molecular data of \citet{abgrall2000}.
The
molecular hydrogen transmission values have been checked against those of
\citet{mcc2003}. Most of the lines in the $FUSE$ spectrum are due to molecular hydrogen.

The ISM parameters are in Table~6. The H1 column
density is in reasonable agreement with \citet{drew2003}. Although Drew et al. present evidence
for a two-component ISM toward QU Car, we have restricted our simulation to a single component.
The output of the simulation
program is a list of transmission values at the same spectral resolution as the system synthetic
spectrum. 
The product of the ISM transmission function 
and our 0.02\AA-resolution
synthetic spectrum produces a corrected synthetic spectrum.	
We have used the $\dot{M}=6.0{\times}10^{-7}~{M}_{\odot}{\rm yr}^{-1}$, $M_{\rm wd}=1.20M_{\odot}$
case to illustrate inclusion of the ISM simulation.

The ISM--corrected synthetic spectrum line density is so great in the $FUSE$ region that it is not
meaningful to present a plot comparable to Figure~6 (and a color plot is not helpful).
Figure~17 shows a partial plot of the $FUSE$ comparison;
note the cutoff at 912\AA~as compared with Figure~6. 
The ISM lines in our new model fit the observed lines well in many cases, including line depths.
There are depth discrepancies in some cases, and our model misses some observed lines.
Our ISM model is a reasonable first order representation of a very complex ISM.
There now is a good fit at Ly~${\gamma}$
and Ly~${\delta}$. Large unmodeled residuals remain from N~IV and S~VI (see Figure~1). 
\citet{har2002} discuss features like these in terms of a wind and/or a disk chromosphere. No
adjustment of parameters in our present set of models can produce a fit to these features,
and they represent the largest remaining residuals.
Note the differences in the ordinates
in Figure~17 and Figure~1. The latter apply to unreddened fluxes and the former to fluxes corrected
for reddening.

Figure~18 shows the fit near Ly~$\beta$. Note the unmodeled broad O VI features at 1033\AA~and 
1037\AA~and the improved Ly~$\beta$ fit as compared with Figure~6. Figure~19 shows the fit to the $STIS$
spectrum. The fit to Ly~$\alpha$ now is accurate (compare with Figure~7). The $STIS$ spectrum
shows a number of unmodeled absorption features including the N~V line pair near 1243\AA~and the
series of absorption lines starting near 1370\AA.

Although we have illustrated correction for the ISM for a single empirical model, the procedure is
applicable to all of those models. 
Our ISM model fit to the
C I lines in the $FUSE$ spectrum does not indicate a C overabundance in the ISM.
\citet{drew2003} present evidence of a C overabundance in the matter undergoing transfer
from the secondary, based on
C emission lines.

\section{Determination of system parameters}

Several conclusions follow from the analysis to this point.

(1) No standard model that we studied fits the observed spectra in either of the two $M_{\rm wd}$ cases.
For reasons discussed previously, we believe the same conclusion would be true
for any intermediate case, and we believe the two cases are limiting cases.

(2) No truncated standard model that we studied produces an acceptable fit to observed spectra in either of 
the $M_{\rm wd}$ cases.
Again, for reasons already discussed, we believe that situation would be true for any intermediate case, 
and that the two cases are
limiting cases.

(3)	Possible $\dot{M}$ values are bounded for small values by an unstable accretion disk. This limit is about
$\dot{M}=1.0{\times}10^{-7}~{M}_{\odot}{\rm yr}^{-1}$.

(4) A possible constraint at high $\dot{M}$ is theoretical: expansion of the WD to
a red giant and consequent engulfing of the entire binary \citep{pacz1978,sion1979}. However, the
studies cited considered spherical accretion and do not apply to disk accretion \citep{pacz1978}.
The effects of mass transfer onto WDs via accretion disks
have been studied by \citet{piro2004}, based on a spreading layer
model. These authors consider $\dot{M}$ values up to $1.5{\times}10^{-7}M_{\odot}{\rm yr}^{-1}$ and
find no significant expansion of the WD. Note that their study considers the mass actually transferred 
to the WD, and in the case of QU Car there is a wind which may carry away an appreciable mass \citep{har2002}.
At $\dot{M}=1.0{\times}10^{-6}~{M}_{\odot}{\rm yr}^{-1}$ and larger
\citet{piro2004} point out that
hydrogen is expected to be burning steadily on the WD surface, leading to luminosities in the range
$10^{37}-10^{38} {\rm erg~s^{-1}}$. A luminosity that large exceeds the disk luminosity and so would
be incompatible with the observed double $\lambda4686$ emission line.
We take $\dot{M}=1.0{\times}10^{-6}~{M}_{\odot}{\rm yr}^{-1}$ to be an upper limit to acceptable
mass transfer rates.
 
(5) Non-standard models can be found that produce acceptable fits to the combined observed
spectra. We achieved successful fits with a hot accretion disk region bracketed between the WD and a radius 
$R=1.3611R_{\rm wd}$ (which we designate as a boundary layer),
an intermediate temperature region adjacent to the hot region, and an isothermal region extending to the
tidal cutoff boundary. The hot region $T_{\rm eff}$ was set equal to the standard model temperature maximum to
preserve a designation connection with the standard model $\dot{M}$. Because the peak emissions of both the
hot region and the intermediate temperature region are at much shorter wavelengths than the short
wavelength limit of the $FUSE$ spectrum, we have no good way to impose tight temperature constraints on those regions.

Table~7 compares properties of models that fit the observed spectra well. 
Columns 3,4,5 list the $T_{\rm eff}$ values for the boundary layer, the intermediate region, and the
isothermal region. The last four lines list the possible range of isothermal region $T_{\rm eff}$
values. The first four lines, for $M_{\rm wd}=1.2M_{\odot}$, present, first, a comparison of line~1 with line~6 for
the same $T_{\rm eff}$ values of the intermediate and isothermal regions.
Line~6, for $M_{\rm wd}=0.6M_{\odot}$, corresponds to a slightly smaller accretion disk (compare $r_a$ in 
Table~4 and Table~5) and the value of $i$ for Table~4 is $40\arcdeg$ versus $55\arcdeg$ for Table~5.
Line~3 and line~4 preserve the same
value of isothermal region $T_{\rm eff}$ and illustrate the increase of bolometric luminosity with,
primarily, the change in the boundary layer $T_{\rm eff}$. Note that the calculated distance to
QU Car is the same for these two lines, indicating that the normalizing factor to fit the
synthetic spectrum to the observed spectra depends strongly on the isothermal region $T_{\rm eff}$ and
is fairly insensitive to the boundary layer $T_{\rm eff}$. 
Column~6 lists the bolometric luminosity of the standard model for comparison
with the bolometric luminosity of the empirical model in
column~7. Note that the entries in the two columns are roughly comparable, in agreement with our earlier
discussion on preserving a connection between the empirical model and the standard model $\dot{M}$.
The last column lists the distance
to QU Car for each empirical model.
The range of distances is not large for $M_{\rm wd}=1.2M_{\odot}$, and all are in rough accord with
the Hipparcos distance (\S2). The smallest distance for $M_{\rm wd}=0.6M_{\odot}$ (332 pc) is in
marginal disagreement with the Hipparcos distance.

There is no clear distinction, among 
the acceptable models, of the quality
of fit. It is important to note that the $FUSE$, $STIS$ and $IUE$ spectra were obtained at different times,
that they were fitted together by visual estimate, and that the QU Car luminosity varies on a short time scale 
\citep{schild1969,kn1994}. Because of this continuing variation, 
a new study with new data likely would find larger differences from the present data than the residuals
in our present model fits. Although accurate fits in some spectral regions are possible (e.g., Figure~8),
the fit accuracy should not be overinterpreted.

The fit
in the 920\AA~to 1200\AA~ region is fairly sensitive to the $T_{\rm eff}$ of the intermediate temperature
region. The width of that region arbitrarily was set to two (BINSYN) annuli, and their corresponding widths 
depend
on the choice of 45 annulus divisions of the accretion disk in the BINSYN model. 
By eye estimate, the fits with the present models are optimal.
We suspect that a choice of one of the models over another, based on, say, a
${\chi}^2$ test, could arise from numerical effects (interpolation approximations, etc.) rather
than from a closer representation of physical reality. Since the largest remaining residuals arise from 
unmodeled wind or chromospheric features, these models present the closest representation
of the QU Car data currently available.

Table~4 and Table~5 list the orbital parameters used for the two values of $M_{\rm wd}$.
Since there is no light curve to simulate, the $T_{\rm eff}$
of the secondary star is of importance only to the extent that it could affect the long wavelength
end of the system synthetic spectrum. A nominal 6500K secondary in the $M_{\rm wd}=1.2M_{\odot}$ system
provides no evidence of contamination in the system synthetic spectrum. Initial retention of that
$T_{\rm eff}$ in the $M_{\rm wd}=0.6M_{\odot}$ system, followed by substitution of a 3500K secondary
produces no detectable effect on the system synthetic spectrum.

We are confident that $M_{\rm wd}$ masses could be selected between our limiting values of $0.6M_{\odot}$
and $1.2M_{\odot}$, combined with appropriate values of $M_{\rm s}$ and $i$ from Figure~2, assigned
a range of $\dot{M}$ values and produce SED fits to the observed spectra comparable in quality to Figure~5 or
Figure~12. Table~7 represents the extremes of possible system models, based on existing spectroscopic data.
Because of its consistency with the empirical mass-period relation \citep[eq.~2.100]{w1995} we favor
the $M_{\rm wd}=1.2M_{\odot}$ solution.

\section{Discussion}

We have appropriated the term "boundary layer`` to designate the region between the WD and
the accretion disk $R=1.3611R_{\rm wd}$. This region more generally is associated with a very hot
region occupying the same space but predicted to emit hard X-rays if optically thin or
soft X-rays if optically thick \citep[sect. 2.5.4]{w1995}. This prediction is modified if
a wind carries off part of the accretion mass \citep{HD1991,HD1993}. In the latter case, predicted
boundary layer $T_{\rm eff}$'s are no higher than the values adopted in our models.	Based both on
its geometric location and the much higher local temperature than in the rest of the accretion
disk, our appropriated term actually is in keeping with the usual sense of the term.

Adoption of the intermediate temperature layer obviously is a crude approximation. A temperature
discontinuity at its inner and outer boundary, as in our present models, is not physically credible. 
However, this model is 
a much better fit to
observation than with our earlier experimental (\S6) models, which adopted a sharp, nonetheless more gradual
temperature drop to the isothermal disk region than in this paper. 

The outermost accretion disk region contibutes little to the system synthetic spectrum. The $T_{\rm eff}$
values, for a limited region which would vary from one $\dot{M}$ to another, could be closer 
to standard model values than the isothermal assumption without
producing a detectable change in the system synthetic spectrum.
On the other hand it is known that temperatures in the outer region of an accretion disk need an
upward correction from the standard model to account for impact heating \citep{las2001} and tidal
heating \citep{bm2001}.

The accretion disk temperature profiles found in this study differ from profiles found for other NL
systems. In both SDSSJ0809 \citep{linnell2007a}  and IX Vel \citep{linnell2007b} the final model included
an inner isothermal region with an outer region that followed a standard model. In the case of MV Lyr 
\citep{l2005} in
a high state, the final model had a standard model accretion disk extending from an inner truncation
radius to an intermediate radius with an isothermal radius beyond. It is of interest that all of the
NL systems included in these studies have accretion disks whose temperature profiles clearly differ from the
standard model.

\section{Summary}

The analysis in this study leads to the following conclusions:

(1) Standard model SEDs, within the tested ranges of $\dot{M}$ and $M_{\rm wd}$,
are too hot to fit 
the combined set of $FUSE$, $STIS$, and $IUE$ spectra of QU Car with acceptable accuracy.

(2) Truncated standard model SEDs fail to fit the combined set of $FUSE$, $STIS$, and $IUE$ spectra of QU Car
for similar reasons.

(3) Alternative non-standard models have SEDs that accurately fit the set of observed spectra. These models
individually have a boundary layer $T_{\rm eff}$ equal to the maximum accretion disk $T_{\rm eff}$ of an 
associated standard model $\dot{M}$. A narrow adjustable $T_{\rm eff}$ annulus is needed adjacent to the
boundary layer to "fine tune`` the SED fit to the $FUSE$ spectrum. The remainder of the accretion disk is
isothermal to within the accuracy of the SED fit. The model-to-model residuals are so similar that it is
not possible to choose a best model. 
The QU Car WD mass is greater than $0.6M_{\odot}$ and
less than the Chandrasekhar limit; a solution with $M_{\rm wd}=1.2M_{\odot}$ is consistent with an
empirical mass-period relation.
The $\dot{M}$ from the secondary star is between 
$1.0{\times}10^{-7}M_{\odot}~{\rm yr}^{-1}$
and $1.0{\times}10^{-6}M_{\odot}~{\rm yr}^{-1}$.
The set of models agree on a limited range
of possible isothermal region $T_{\rm eff}$ values between 14,000K and 18,000K.

The orbital inclination is between $i=40\arcdeg$ and $i=60\arcdeg$. 
The Hipparcos distance, 610 pc, is representative of the empirical model results.

(4) A model of the ISM
combines with the empirical accretion disk models to provide an improved fit to the observed $FUSE$ and 
$STIS$ spectra.	The largest remaining residuals are from unmodeled, broad, high excitation features
probably associated with a wind or a disk chromosphere.

(5) The accretion disk dominates the system flux in the $FUSE$, $STIS$, and $IUE$ spectral regions.
An assumed 55,000K WD provides a negligible flux contribution, as does the accretion disk rim and 
the secondary star.

The authors are grateful to the referee for a careful and prompt consideration of the initial version
of this paper. The final version has benefitted substantially from the criticisms of the referee.
PG is thankful to Mario Livio for his kind hospitality at the Space Telescope Science Institute.
Support for this work was provided by NASA through grant number 
HST-AR-10657.01-A  to Villanova University (P. Godon) from the Space
Telescope Science Institute, which is operated by the Association of
Universities for Research in Astronomy, Incorporated, under NASA
contact NAS5-26555. 
PS is supported by HST grant GO-09724.06A.

This research was partly based on observations made with the NASA/ESA Hubble Space Telescope, obtained at the
Space Telescope Science Institute, which is operated by the Association of Universities 
for Research in Astronomy, Inc. under NASA contract NAS5-26555, and the NASA-CNES-CSA
{\it Far Ultraviolet Explorer}, which is operated for NASA by the Johns Hopkins University
under NASA contract NAS5-32985.

\clearpage



\begin{deluxetable}{lllcl}
\tablewidth{0pt}
\tablenum{1}
\tablecaption{FUSE Lines
}
\tablehead{
\colhead{Wavelength}  &  \colhead{Line} & \colhead{Origin} 
& \colhead{FWHM}	& \colhead{Comments}\\
\colhead{(\AA)}	&  \colhead{Identification} &  & \colhead{(\AA)}&
}
\startdata
see text   &  H\,{\sc i}                &   c,ism &         &   \\
see text   &  H\,{\sc ii}               &   c,ism  &        &   \\
see text   &  D\,{\sc i}                &   c,ism  &        &   \\
see text   &  O\,{\sc i}                &   c,ism  &        &   \\
923.0      &  N\,{\sc iv} 921.46-924.91 &    s      & 4.0 & unresolved multiplet \\ 
933.3      &  S\,{\sc vi} 933.38        &    s      & 3.0 & affected by $H_2$  \\
944.6      &  S\,{\sc vi} 944.70        &    s      & 2.5 & affected by $H_2$    \\
950.5      &  P\,{\sc iv} 950.66        &    s?     & & heavily affected by $H_2$\\ 
951.05     &  N\,{\sc i}  951.12        &   c,ism  &  0.1 &     \\
953.35     &  N\,{\sc I}  953.42        &   c,ism  &  0.1 &     \\
953.60     &  N\,{\sc I}  953.66        &   c,ism  &  0.1 &     \\ 
953.97     &  N\,{\sc I} 953.97+954.10  &   c,ism  &  0.3 & unresolved   \\
961.00     &  P\,{\sc ii} 961.04        &   c,ism  &  0.1 &     \\
962.10     &  P\,{\sc ii} 962.12        &   c,ism  &  0.1 &     \\
963.75     &  P\,{\sc ii} 963.80        &   c,ism  &  0.1 & affected by $H_2$\\ 
963.95     &  N\,{\sc i}  963.99        &   c,ism  &  0.1  &    \\
964.57     &  N\,{\sc i}  964.63        &   c,ism  &  0.1   &   \\
977.00     &  C\,{\sc iii} 977.03        &     c,ism  & 0.15 &      \\
977.00     &  C\,{\sc iii} 977.03        &  s?        & 2.0 & affected by $H_2$\\ 
989.80     &  N\,{\sc iii} 989.77+Si\,{\sc ii} 989.89 &  c,ism & & affected by $H_2$ \\
1031.80    &  O\,{\sc vi}  1031.65       &   s       &  3.5   &   \\
1036.34    &  C\,{\sc ii}  1036.34~doublet      &   ism     &  & affected by $H_2$\\  
1037.2?    &  0\,{\sc vi}  1037.30       &   s       &  3.0 & heavily affected by $H_2$\\ 
1048.20    &  Ar\,{\sc i}  1048.22       &   ism     &  0.1  &    \\ 
1055.25    &  Fe\,{\sc ii} 1055.27       &   ism     &  0.1   &   \\ 
1062.6?    &  S\,{\sc iv}  1062.66       &    s      &  & heavily affected by $H_2$\\ 
1063.15    &  Fe\,{\sc ii} 1063.18       &   ism     &  0.1 &     \\ 
1066.65    &  Ar\,{\sc i}  1066.66       &   ism     &  0.1  &    \\ 
1073.2     &  S\,{\sc iv}  1073.1+1073.5 &    s      & 2.0 & unresolved doublet \\ 
1081.85    &  Fe\,{\sc ii} 1081.87       &   c,ism   & & affected by $H_2$\\ 
1083.97    &  N\,{\sc ii}  1083.99       &   c,ism   & 0.15 &      \\
1096.85    &  Fe\,{\sc ii} 1096.88       &   c,ism   & 0.1 & affected by $H_2$\\ 
1112.00    &  Fe\,{\sc ii} 1112.03       &   c,ism   & 0.1  &      \\ 
1118.05    &  S\,{\sc iv, vi} 1117.9+P\,{\sc v} 1117.98 &  s  & 1.7 &        \\ 
1121.97    &  Fe\,{\sc ii} 1121.97       &   c,ism   &  0.1 &      \\ 
1122.40    &  Si\,{\sc iv} 1122.48       &   s       &  1.5  &    \\ 
1128.20    &  P\,{\sc v} 1128.01+Si\,{\sc iv} 1128.33 &  s & 1.5 &   \\
1133.70    &  Fe\,{\sc ii} 1133.67       &   c,ism   &  0.1  &     \\ 
1134.15    &  N\,{\sc i}   1134.17       & a,c,ism   &  0.1   &    \\
1134.45    &  N\,{\sc i}   1134.42       & a,c,ism   &  0.1  &     \\
1135.00    &  N\,{\sc i}   1134.98       & a,c,ism   &  0.15  &    \\
1142.40    &  Fe\,{\sc ii} 1142.37       &  c,ism    &  0.1    &   \\
1143.25    &  Fe\,{\sc ii} 1143.23       &  c,ism    &  0.1 &      \\
1144.95    &  Fe\,{\sc ii} 1144.94       &  c,ism    &  0.1  &     \\
1152.85    &  P\,{\sc ii}  1152.82       &  c,ism    &  0.1   &    \\
1168.75    &  He\,{\sc i} 1168.9         &  a?s?     & 0.3 & emission \\  
1172.50    &  S\,{\sc i} 1172.55?        &  c,ism    & 0.3 &  \\  
1173.00    &  O\,{\sc iii} 1173.37       &  s?       & 0.3 &   \\ 
1175.95    &  C\,{\sc iii} 1174.9-1176.4 &   s       & 1.7 & shifted by +0.35\AA    \\ 
\enddata
\tablecomments{a=possibly including air glow contamination; c=circumbinary material; 
ism=ISM; s=source}
\end{deluxetable}

\clearpage

\begin{deluxetable}{lllcl}
\tablewidth{0pt}
\tablenum{2}
\tablecaption{STIS Lines
}
\tablehead{
\colhead{Wavelength}  &  \colhead{Line} & \colhead{Origin} 
& \colhead{FWHM}	& \colhead{Comments}\\
\colhead{(\AA)}	&  \colhead{Identification} &  & \colhead{(\AA)}&
}
\startdata
1168.59 & N\,{\sc i} 1168.54       &  c,ism      &  0.1    &   \\ 
1171.08 & N\,{\sc i} 1171.08       &  c,ism      &  0.15   &   \\ 
1175.7  & C\,{\sc iii}             & s           &         & unresolved multiplet  \\ 
1188.83 & C\,{\sc i} 1188.83       &  c,ism      &  0.1    &   \\ 
1190.23 & Si\,{\sc ii} 1190.20     &  c,ism      & 0.07    &   \\
1190.29 & C\,{\sc i} 1190.25       &  c,ism      &         & unresolved  \\ 
1190.43 & Si\,{\sc ii} 1190.42     &  c,ism      & 0.13    & \\
1193.05 & C\,{\sc i} 1193.03       &  c,ism      &  0.07   &   \\ 
1193.30 & Si\,{\sc ii} 1193.29     &  c,ism      & 0.16    &   \\
1197.21 & Mn\,{\sc ii} 1197.18     &  c,ism      & 0.05    &   \\
1199.43 & Mn\,{\sc ii} 1199.39     &  c,ism      & 0.05    &   \\
1199.56 & N\,{\sc i} 1199.55       &  c,ism      & 0.16    &   \\
1200.23 & N\,{\sc i} 1200.22       &  c,ism      & 0.14    &   \\
1200.72 & N\,{\sc i} 1200.71       &  c,ism      & 0.13    &   \\
1201.15 & Mn\,{\sc ii} 1201.12     &  c,ism      & 0.05    &   \\
1206.53 & Si\,{\sc iii} 1206.50    &  c,ism      & 0.12    &   \\
1215.7  & H\,{\sc i} 1215.66    &  s,c,ism      & 12      &   \\
1237.09 & Ge\,{\sc ii} 1237.06     &  c,ism      & 0.06    &   \\
1238.94 & N\,{\sc v} 1238.82     &  s          & 2.2     &   \\
1239.96 & Mg\,{\sc ii} 1239.93     &  c,ism      & 0.05    &   \\
1240.42 & Mg\,{\sc ii} 1240.39     &  c,ism      & 0.05    &   \\
1243.00 & N\,{\sc v} 1242.80     &  s          & 1.7     &   \\
1250.62 & S\,{\sc ii} 1250.58      &  c,ism      & 0.11    &   \\
1253.83 & S\,{\sc ii} 1253.81      &  c,ism      & 0.12    &   \\
1259.53 & S\,{\sc ii} 1259.52      &  c,ism      & 0.12    &   \\
1260.45 & Si\,{\sc ii} 1260.42?    &    ?        & 0.26    & asymmetric     \\
1260.77 & C\,{\sc i} 1260.74       &  c,ism      & 0.06    &   \\
1266.4  & C\,{\sc i} 1266.42       &  s          & 1.6     & shallow \\ 
1272.7  & O\,{\sc ii} 1272.23       &  s          & 1.5     & shallow \\ 
1277.29 & Zn\,{\sc ii} 1277.31?    &  c,ism      & 0.08    &   \\
        & C\,{\sc i} 1277.25+1277.28?&  c,ism      & 0.08  &   \\
1277.57?& C\,{\sc i} 1277.51+1277.55?&  c,ism      & 0.12  & very shallow, unresolved  \\
1280.18 & C\,{\sc i} 1280.14       &  c,ism      & 0.05    &  shallow     \\
1301.91 & P\,{\sc ii} 1301.87      &  c,ism      & 0.05    &    \\
1302.19 & O\,{\sc i} 1302.17       &  c,ism      & 0.20    &      \\ 
1304.39 & Si\,{\sc ii} 1304.37     &  c,ism      & 0.14    &      \\ 
1304.85 & O\,{\sc i} 1304.86       &  c,ism      & 0.20    & very shallow     \\ 
1306.02 & O\,{\sc i} 1306.03       &  c,ism      & 0.20    & very shallow     \\ 
1317.25 & Ni\,{\sc ii} 1317.22     &  c,ism      & 0.09    &  shallow     \\ 
1328.87 & C\,{\sc i} 1328.83       &  c,ism      & 0.06    &  shallow     \\ 
1329.14 & C\,{\sc i} 1329.12       &  c,ism      & 0.06    &  very shallow     \\ 
1334.56 & C\,{\sc ii} 1334.53      &  c,ism      & 0.21    &       \\ 
1335.73 & C\,{\sc ii} 1335.71      &  c,ism      & 0.14    &       \\ 
1347.28 & Cl\,{\sc i} 1347.24      &  c,ism      & 0.05    &  shallow     \\ 
1355.65 & O\,{\sc i} 1355.60       &  c,ism      & 0.05    &  very shallow     \\ 
1358.81 & Cu\,{\sc ii} 1358.77     &  c,ism      & 0.05    &  very shallow     \\ 
1370.17 & Ni\,{\sc ii} 1370.14     &  c,ism      & 0.07    &  shallow     \\ 
1371.6  & O\,{\sc v} 1371.30       &  s          & 1.6     &               \\  
1393.95 & Si\,{\sc iv} 1393.76     &  s          & 1.1     &     \\
1403.95 & Si\,{\sc iv} 1402.77     &  s          & 1.2     &     \\ 
1420.7  & ?                        &  s?         & 1.7     & seen only in one exposure \\ 
1414.43 & Ga\,{\sc ii} 1414.40     &  c,ism      & 0.05    &  very shallow     \\ 
1454.89 & Ni\,{\sc ii} 1454.84     &  c,ism      & 0.04    &  very shallow     \\ 
1526.73 & Si\,{\sc ii} 1526.71     &  c,ism      & 0.17    &                   \\ 
1532.58 & P\,{\sc ii} 1532.53      &  c,ism      & 0.05    &  very shallow     \\ 
1548.18 & C\,{\sc iv} 1548.19      &  s          & 1.0     &     \\ 
1550.85 & C\,{\sc iv} 1550.77      &  s          & 1.0     &     \\ 
1560.36 & C\,{\sc i} 1560.31       &  c,ism      & 0.06    &                   \\ 
1560.74 & C\,{\sc i} 1560.71       &  c,ism      & 0.07    & very shallow      \\ 
1640.52 & He\,{\sc ii} 1640.47     &  s          & 1.0     &    \\ 
1656.98 & C\,{\sc i} 1656.93       &  c,ism      & 0.07    &                   \\ 
1657.43 & C\,{\sc i} 1657.38       &  c,ism      & 0.04    &  very shallow   \\ 
1657.96 & C\,{\sc i} 1657.91       &  c,ism      & 0.04    &  very shallow  \\ 
1670.81 & Al\,{\sc i} 1670.79      &  c,ism      & 0.13    &    \\ 
\enddata
\tablecomments{a=possibly including air glow contamination; c=circumbinary material; 
ism=ISM; s=source}
\end{deluxetable}


\begin{deluxetable}{rrrrrrr}
\tablewidth{0pt}
\tablenum{3}
\tablecaption{Properties of accretion disk with mass transfer rate 
$\dot{M}=6.0{\times}10^{-7}~{M}_{\odot}{\rm yr}^{-1}$ and WD mass of $1.20{M}_{\odot}$.}
\tablehead{	  
\colhead{$r/r_{\rm wd,0}$} & \colhead{$T_{\rm eff}$} & \colhead{$m_0$} 
& \colhead{$T_c$} & \colhead{log~$g$}
& \colhead{$z_H$} & \colhead{{$\tau_{\rm Ross}$}}}
\startdata
1.36  &  334084  &  2.213E5  & 4.534E6  &  7.88   & 6.92E7   & 8.73E4\\
2.00  &	 299540  &  2.701E5	 & 4.221E6  &  7.62   & 1.21E8   & 1.02E5\\
3.00  &	 242217  &  2.640E5	 & 3.386E6  &  7.31   & 1.99E8   & 9.93E4\\
4.00  &	 203586  &  2.458E5	 & 2.802E6  &  7.08   & 2.76E8   & 9.34E4\\
5.00  &	 176588  &  2.284E5	 & 2.394E6  &  6.89   & 3.56E8   & 8.80E4\\
6.00  &	 156665  &  2.133E5	 & 2.097E6  &  6.75   & 4.38E8   & 8.36E4\\
7.00  &	 141312  &  2.004E5	 & 1.871E6  &  6.62   & 5.20E8   & 7.99E4\\
8.00  &	 129082  &  1.894E5	 & 1.693E6  &  6.51   & 6.04E8   & 7.68E4\\
9.00  &	 119082  &  1.798E5	 & 1.550E6  &  6.42   & 6.90E8   & 7.43E4\\
10.00 &	 110733  &  1.714E5	 & 1.432E6  &  6.33   & 7.77E8   & 7.22E4\\
12.00 &	 97540   &  1.572E5	 & 1.247E6  &  6.18   & 9.54E8   & 6.88E4\\
14.00 &	 87537   &  1.460E5	 & 1.110E6  &  6.06   & 1.14E9   & 6.63E4\\
16.00 &	 79657   &  1.366E5	 & 1.004E6  &  5.95   & 1.32E9   & 6.45E4\\
18.00 &	 73268   &  1.288E5	 & 9.197E5  &  5.85   & 1.51E9   & 6.31E4\\
20.00 &	 67967   &  1.221E5	 & 8.505E5  &  5.77   & 1.71E9   & 6.21E4\\
24.00 &	 59649   &  1.110E5	 & 7.439E5  &  5.62   & 2.11E9   & 6.08E4\\
30.00 &	 53387   &  1.024E5	 & 6.334E5  &  5.50   & 2.73E9   & 6.02E4\\
35.00 &	 45436   &  9.088E4	 & 5.682E5  &  5.32   & 3.27E9   & 6.03E4\\
45.00 &	 37854   &  7.930E4	 & 4.782E5  &  5.12   & 4.37E9   & 6.20E4\\
60.00 &	 30685   &  6.767E4	 & 3.957E5  &  4.89   & 6.11E9   & 6.61E4\\
80.00 &  24852   &  5.762E4  & 3.304E5  &  4.66   & 8.53E9   & 7.35E4\\
100.00 & 21092   &  5.081E4	 & 2.891E5  &  4.49   & 1.10E10  & 8.21E4\\
120.00 & 18440   &  4.581E4	 & 2.601E5  &  4.34   & 1.37E10  & 9.15E4\\
140.00 & 16458   &  4.195E4	 & 2.383E5  &  4.22   & 1.64E10  & 1.02E5\\
160.00 & 14911   &  3.885E4	 & 2.211E5  &  4.11   & 1.91E10  & 1.12E5\\
180.00 & 13667   &  3.631E4	 & 2.069E5  &  4.01   & 2.19E10  & 1.22E5\\
\enddata
\tablecomments{Each line in the table represents a separate annulus.
A \citet{ss1973} viscosity parameter $\alpha=0.1$ was used in calculating all annuli.
$r_{\rm wd,0}$ is the radius of a zero temperature, carbon, Hamada-Salpeter WD.
See the text for additional comments.}		 
\end{deluxetable}


\begin{deluxetable}{llll}
\tablewidth{0pt}
\tablenum{4}
\tablecaption{QU Car Model System Parameters, $M_{\rm wd}=1.2M_{\odot}$}
\tablehead{
\colhead{parameter} & \colhead{value} & \colhead{parameter} & \colhead{value}}
\startdata
${ M}_{\rm wd}$  &  $1.20{M}_{\odot}$(adopted)	 &$r_s$(pole) &  $1.185R_{\odot}$  \\
${M}_{\rm s}$  &  $1.2{M}_{\odot}$	         &$r_s$(point)   & $1.664R_{\odot}$\\
P    &  0.4542 days	  & 						$r_s$(side)  & $1.245R_{\odot}$\\
$D$              &  $3.3275R_{\odot}$ & $r_s$back)   & $1.348R_{\odot}$\\	 
${\Omega}_{\rm wd}$         & 500.0 & 	 log $g_s$(pole) & 4.39\\
${\Omega}_s$                &  3.75 & log $g_s$(point) & -5.00\\
{\it i}              &   $40{\degr}$(adopted)& log $g_s$(side)  & 4.30\\
$T_{\rm eff,wd}$         &  $55,000$K(nominal)  &log $g_s$(back)  & 4.15\\
$r_{\rm wd}$      &   $0.00666R_{\odot}$ &		 $r_a$ & $0.998R_{\odot}$ \\
log $g_{\rm wd}$  &   8.9&				  $r_b$ & $0.00666R_{\odot}$ \\
  &               &		  $H$    & $0.05R_{\odot}$  \\

\enddata
\tablecomments{${\rm wd}$ refers to the WD; $s$ refers to the secondary star.
$D$ is the component separation of centers,
${\Omega}$ is a Roche potential. 
$r_a$ specifies the outer radius 
of the accretion disk, set at the tidal cut-off radius, 
and $r_b$ is the accretion disk inner radius. 
$H$ is 
the semi-height of the accretion disk rim (standard model).}  
\end{deluxetable}

\clearpage
\begin{deluxetable}{llll}
\tablewidth{0pt}
\tablenum{5}
\tablecaption{QU Car Model System Parameters, $M_{\rm wd}=0.60M_{\odot}$}
\tablehead{
\colhead{parameter} & \colhead{value} & \colhead{parameter} & \colhead{value}}
\startdata
${ M}_{\rm wd}$  &  $0.60{M}_{\odot}$(adopted)	 &$r_s$(pole) &  $0.8748R_{\odot}$  \\
${M}_{\rm s}$  &  $0.50{M}_{\odot}$	         &$r_s$(point)   & $1.235R_{\odot}$\\
P    &  0.4542 days	  & 						$r_s$(side)  & $0.9169R_{\odot}$\\
$D$              &  $2.56556R_{\odot}$ & $r_s$back)   & $0.9977R_{\odot}$\\	 
${\Omega}_{\rm wd}$         & 162.2 & 	 log $g_s$(pole) & 4.27\\
${\Omega}_s$                &  3.47364 & log $g_s$(point) & 0.43\\
{\it i}              &   $55{\degr}$(adopted)& log $g_s$(side)  & 4.19\\
$T_{\rm eff,wd}$         &  $55,000$K(nominal)  &log $g_s$(back)  & 4.03\\
$r_{\rm wd}$      &   $0.0159R_{\odot}$ &		 $r_a$ & $0.84R_{\odot}$ \\
log $g_{\rm wd}$  &   7.8&				  $r_b$ & $0.0159R_{\odot}$ \\
  &               &		  $H$    & $0.05R_{\odot}$  \\
\enddata
\tablecomments{${\rm wd}$ refers to the WD; $s$ refers to the secondary star.
$D$ is the component separation of centers,
${\Omega}$ is a Roche potential. 
$r_a$ specifies the outer radius 
of the accretion disk, set at the tidal cut-off radius, 
and $r_b$ is the accretion disk inner radius. 
$H$ is 
the semi-height of the accretion disk rim (standard model).}  
\end{deluxetable}

\clearpage

\begin{deluxetable}{ll}
\tablewidth{0pt}
\tablenum{6} 
\tablecaption{Parameters of ISM model}
\tablehead{ 
\colhead{Parameter}           & \colhead{Value}}     
\startdata
N(H)                & $2.5{\times}10^{20} {\rm cm}^{-2}$\\
D/H                 & $1{\times}10^{-5}$\\
N(H2)               & $7{\times}10^{15} {\rm cm}^{-2}$\\
vel.                & $0.0 {\rm~km~s^{-1}}$\\
temp.               & 350 K\\
metallicity         & 0.01 (1.0 = solar)\\
\enddata
\end{deluxetable}

\clearpage
\begin{deluxetable}{rrrrrrrr}
\tablewidth{0pt}
\tablenum{7}
\tablecaption{
Properties, empirical mass transfer models
}
\tablehead{	  
\colhead{$M$} & \colhead{$\dot{M}$} & 
\colhead{$T_{\rm eff}$} 
& \colhead{$T_{\rm eff}$} & \colhead{$T_{\rm eff}$} 
& \colhead{$L_{\rm bol}$}
& \colhead{$L_{\rm bol}$}
& \colhead{$D$}\\
\colhead{($M_{\odot}$)} 
& \colhead{$(M_{\odot}~{\rm yr}^{-1})$} 
& \colhead{(b.l.,K)} 
& \colhead{(inter.,K)} 
& \colhead{(isoth.,K)} 
& \colhead{(std.)} 
& \colhead{(empir.)} 
& \colhead{(pc.)}
}
\startdata
1.20    &  $1.0{\times}10^{-7}$	&  185,000     & 90,000 &   15,000  &   $1.68{\times}10^{36}$ & $1.06{\times}10^{36}$ & 550\\
1.20    &  $3.0{\times}10^{-7}$ &  240,000     &100,000 &   16,000  &   $5.05{\times}10^{36}$ & $2.18{\times}10^{36}$ & 610\\
1.20    &  $6.0{\times}10^{-7}$	&  290,000     &110,000 &   17,000  &   $1.01{\times}10^{37}$ & $3.95{\times}10^{36}$ & 687\\
1.20    &  $1.0{\times}10^{-6}$	&  330,000     &110,000 &   17,000  &   $1.68{\times}10^{37}$ & $5.52{\times}10^{36}$ & 687\\
0.60    &  $1.0{\times}10^{-7}$	&  100,355	   & 70,000	&	14,000	&	$2.56{\times}10^{35}$ & $2.77{\times}10^{35}$ & 332\\
0.60    &  $3.0{\times}10^{-7}$ &  132,000     & 90,000 &   15,000	&   $7.68{\times}10^{35}$ & $7.11{\times}10^{35}$ & 406\\
0.60    &  $6.0{\times}10^{-7}$	&  157,000     &100,000 &   17,000  &   $1.54{\times}10^{36}$ & $1.22{\times}10^{36}$ & 475\\
0.60    &  $1.0{\times}10^{-6}$	&  178,000     &110,000 &   18,000  &   $2.56{\times}10^{36}$ & $1.86{\times}10^{36}$ & 518\\
\enddata
\tablecomments{
b.l.= boundary layer; inter.= intermediate region; isoth.= isothermal region; std.= standard model; empir.= empirical model
The units of columns 6 and 7 are ${\rm erg}~{\rm s}^{-1}$.
}

\end{deluxetable}

\clearpage


\clearpage

\begin{figure}[tb]
\epsscale{0.97}
\plotone{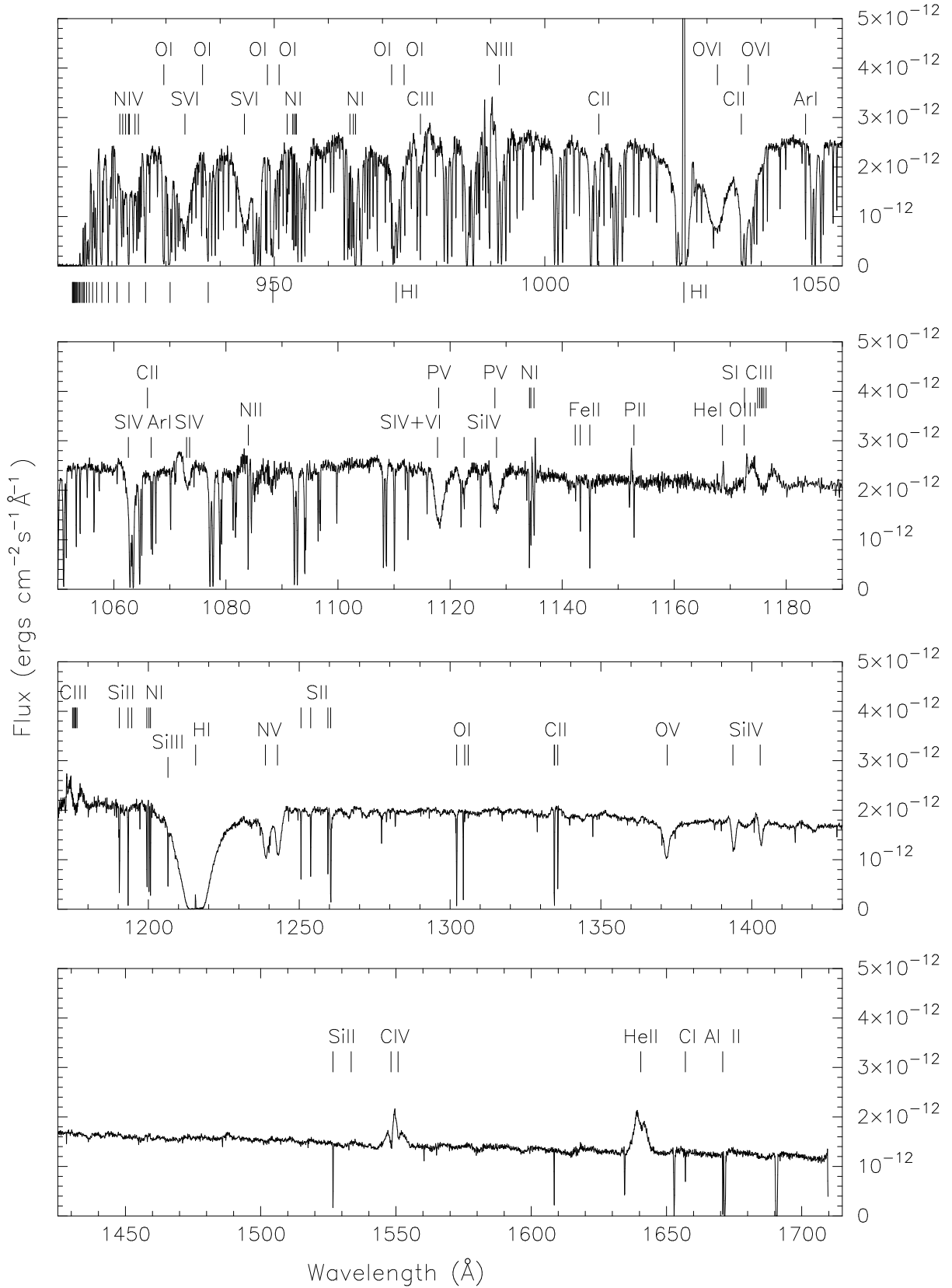}
\epsscale{0.90}
\figcaption{
$FUSE$ plus $STIS$ combined spectra with line identifications. The
$STIS$ spectrum has been multiplied by 1.263 to match the flux level
of the $FUSE$ spectrum in the overlap region.
\label{f1}}
\end{figure}

\begin{figure}[tb]
\epsscale{0.97}
\plotone{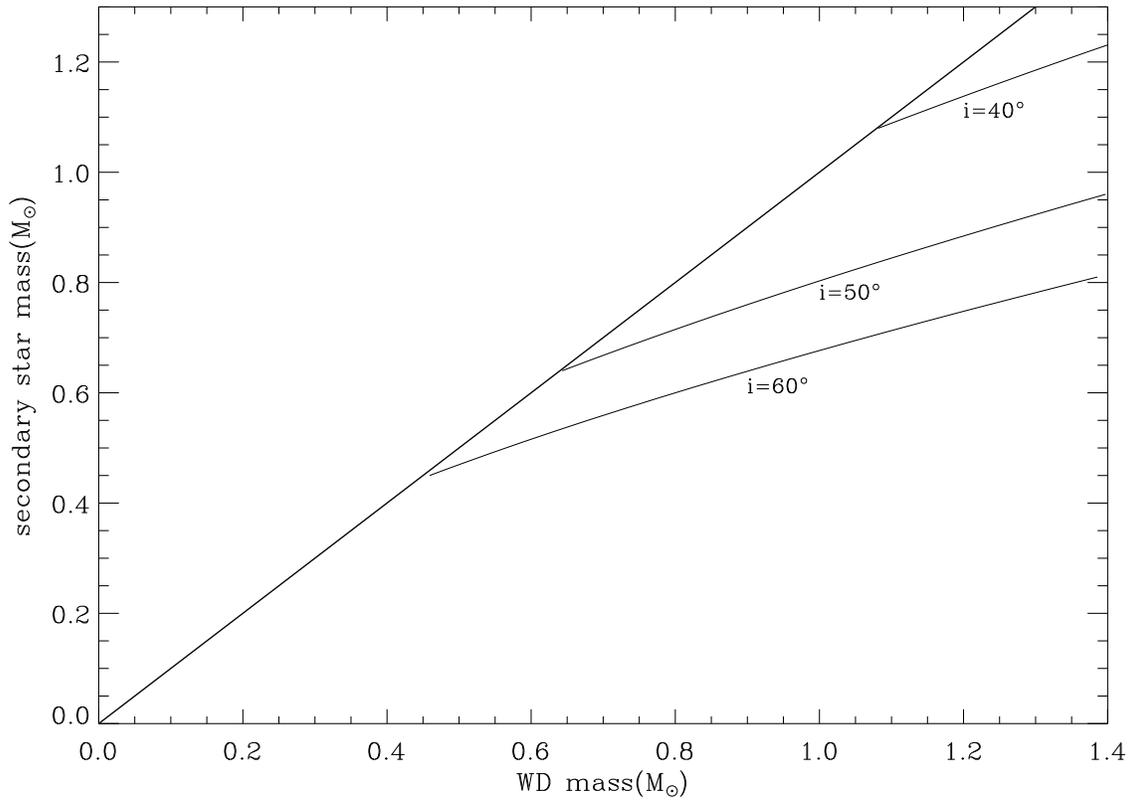}
\epsscale{1.00}
\vspace{0.2cm}
\figcaption{
Mass-mass plot showing regions where a system model is permitted.
The region above the diagonal line is excluded by dynamical instability.
The region below $i=60\arcdeg$ is excluded by the absence of observed
eclipses.
See the text for details.
\label{f2}}
\end{figure}

\begin{figure}[tb]
\epsscale{0.97}
\plotone{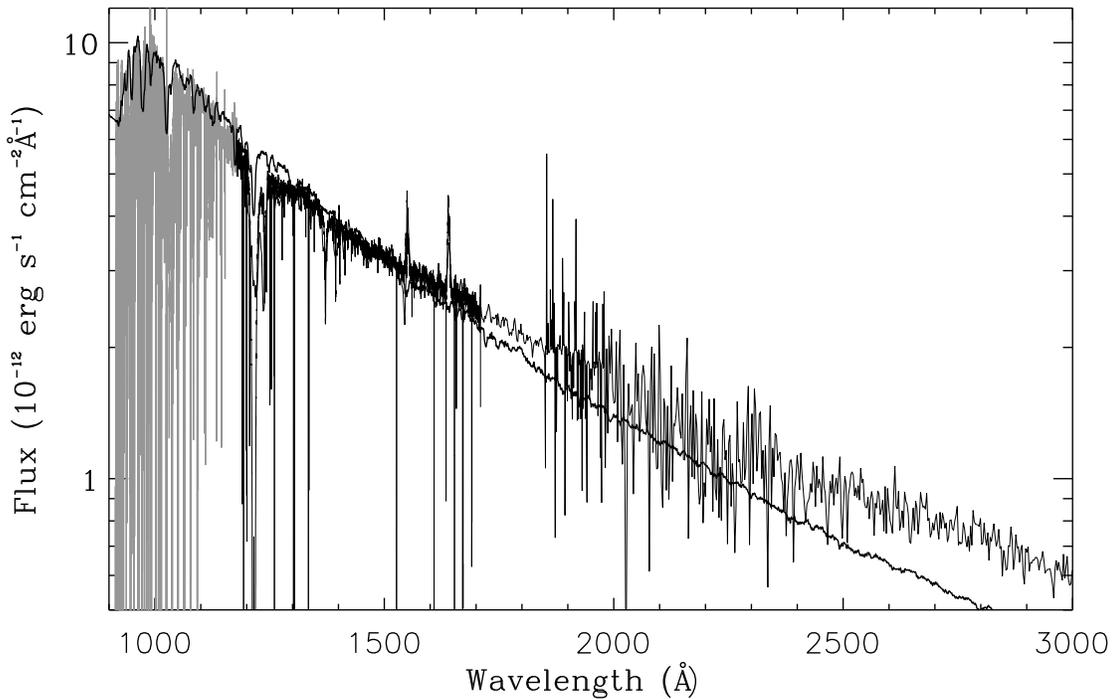}
\epsscale{1.00}
\vspace{0.2cm}
\figcaption{Standard model fit to $FUSE$ + $STIS$ + $IUE$ spectra.
The synthetic spectrum is a continuum spectrum + H and He II lines.
The mass transfer rate is $\dot{M}=6.0{\times}10^{-7}~{M}_{\odot}{\rm yr}^{-1}$
and the WD mass is $1.20M_{\odot}$.
The synthetic spectrum spectral gradient is too large. Note that the
ordinate scale is logarithmic.
\label{f3}}
\end{figure}

\begin{figure}[tb]
\epsscale{0.97}
\plotone{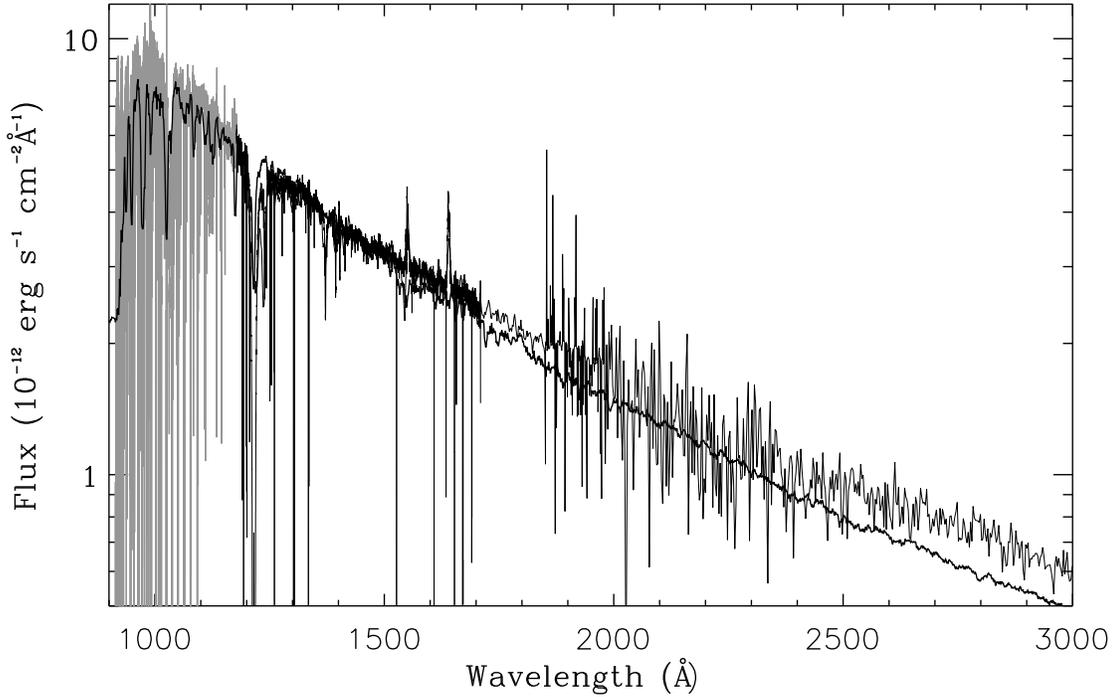}
\epsscale{1.00}
\vspace{0.2cm}
\figcaption{Truncated model fit to observed spectra.
The mass transfer rate is $\dot{M}=6.0{\times}10^{-7}~{M}_{\odot}{\rm yr}^{-1}$,
the WD mass is $1.20M_{\odot}$,
and the accretion disk has been truncated at $R_{\rm trun}=22.5R_{\rm wd}$.
The spectral gradient is a better fit than Figure~3 but still unacceptable
and the synthetic spectrum fails to
fit the FUV accurately.
\label{f4}}
\end{figure}

\clearpage

\begin{figure}[tb]
\epsscale{0.97}
\plotone{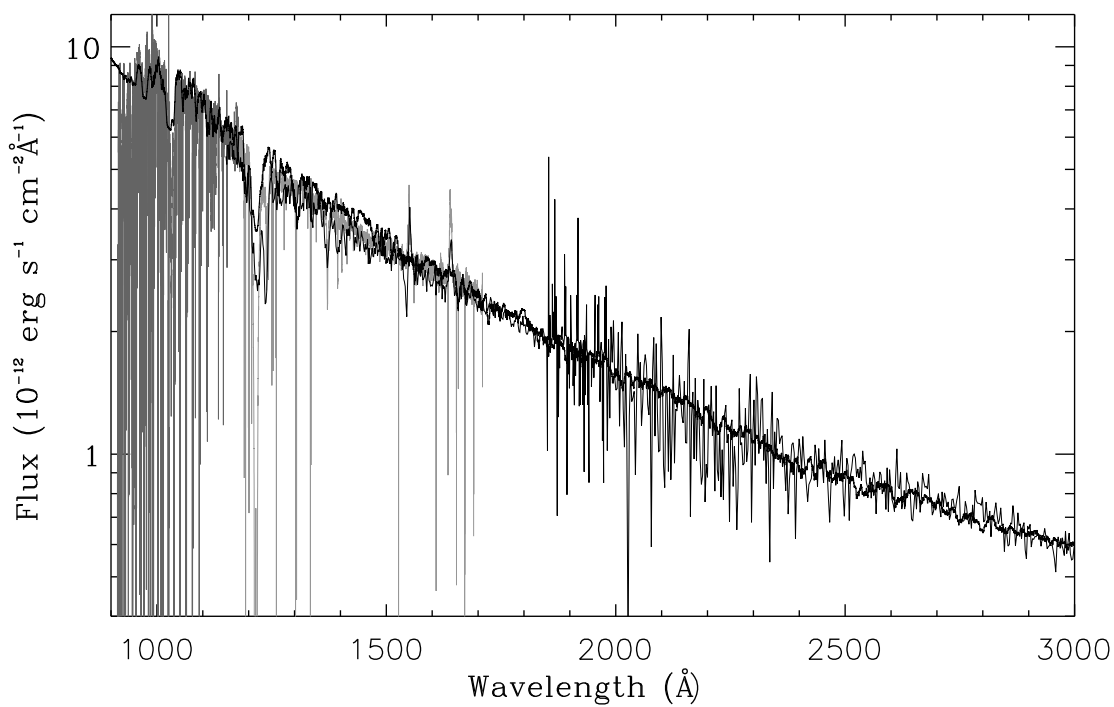}
\epsscale{1.00}\
\vspace{6pt}
\figcaption{Comparison to observed spectra of
model with 17,000K accretion disk face and associated 
$\dot{M}=6.0{\times}10^{-7}~{M}_{\odot}{\rm yr}^{-1}$.
The WD mass is $1.20M_{\odot}$.
The synthetic  (line) spectrum (heavy line) has a resolution of 0.02\AA.
The synthetic spectrum has been divided by $4.5{\times}10^{42}$.
The corresponding distance to QU Car is 687pc. See the text for details.
\label{f5}}
\end{figure}

\begin{figure}[tb]
\epsscale{0.97}
\plotone{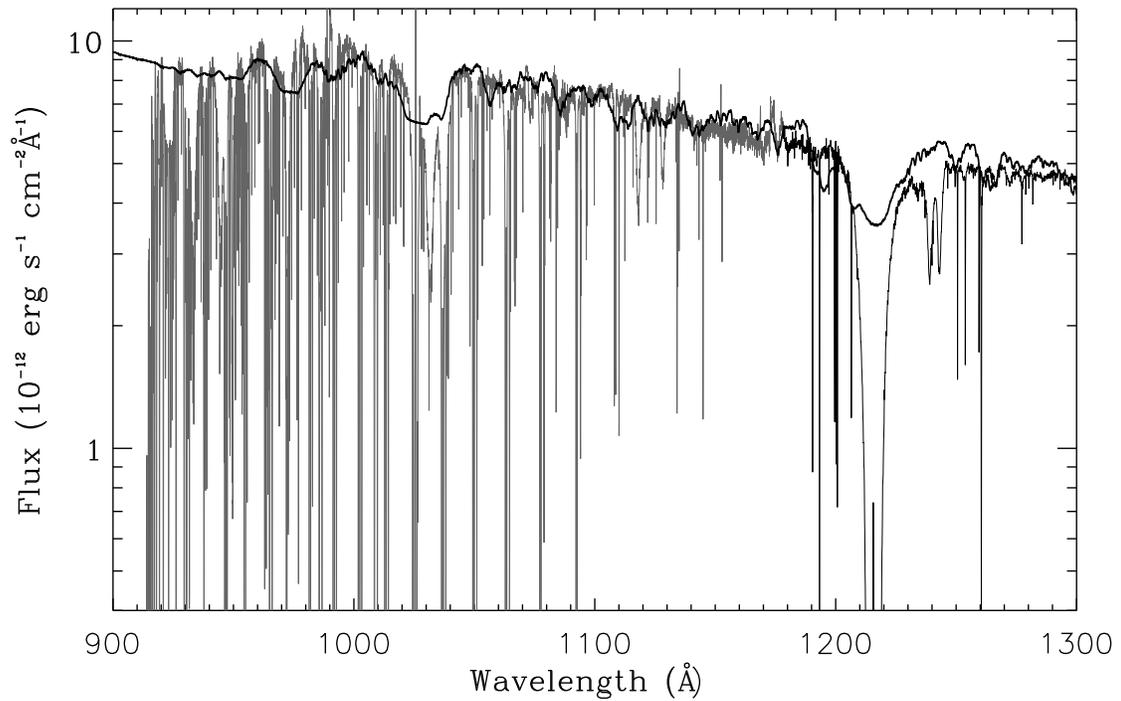}
\epsscale{1.00}
\vspace{6pt}
\figcaption{
Detail of Figure~5. The synthetic spectrum (heavy line) fits the peaks of the $FUSE$
spectrum (grey plot) but the latter spectrum shows strong effects of ISM absorption.
The fit includes the short wavelength end of the $STIS$ spectrum (light line).
See the text for details.
\label{f6}}
\end{figure}

\begin{figure}[tb]
\epsscale{0.97}
\plotone{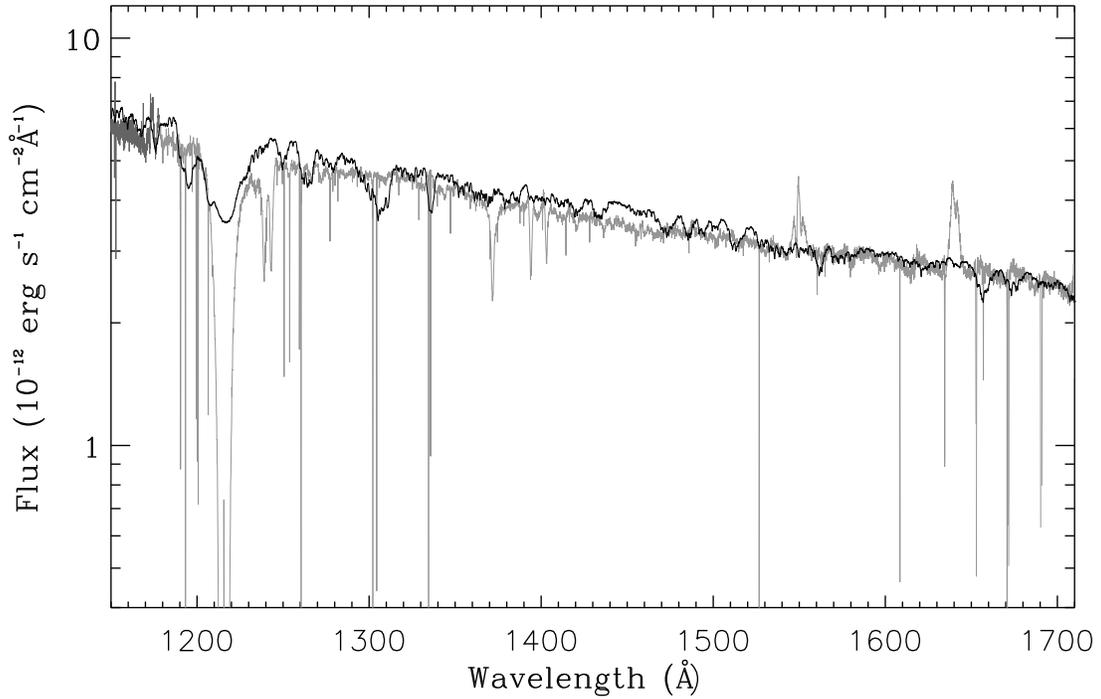}
\epsscale{1.00}
\vspace{6pt}
\figcaption{
Detail of Figure~5 showing fit to the $STIS$ spectrum. ISM absorption 
affects the $STIS$ spectrum shortward
of 1500\AA. 
The $IUE$ contribution has been suppressed to avoid confusion with the $STIS$
spectrum. 
See the text for a discussion. 
\label{f7}}
\end{figure}

\begin{figure}[tb]
\epsscale{0.97}
\plotone{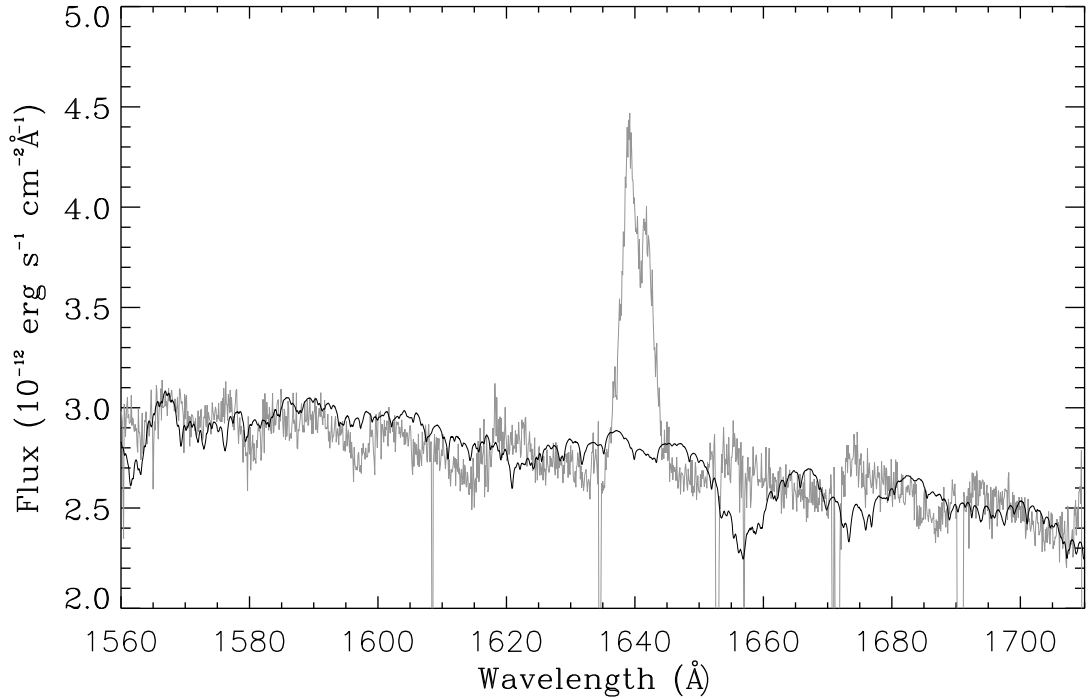}
\epsscale{1.00}
\vspace{6pt}
\figcaption{
Expanded scale detail of Figure~5, synthetic spectrum resolution 0.02\AA. 
The grey plot is a segment of the $STIS$ spectrum, resolution 0.1\AA.
Note that the ordinate scale is linear.
Except for a few details and the He II emission line, the fit is quite accurate.
Note the doubling of the He~II line. The synthetic spectrum normalizing 
factor is the same as in Figure~5 (i.e., the normalizing factor has not been 
changed to produce a better fit in this spectral region).
\label{f8}}
\end{figure}

\begin{figure}[tb]
\epsscale{0.97}
\plotone{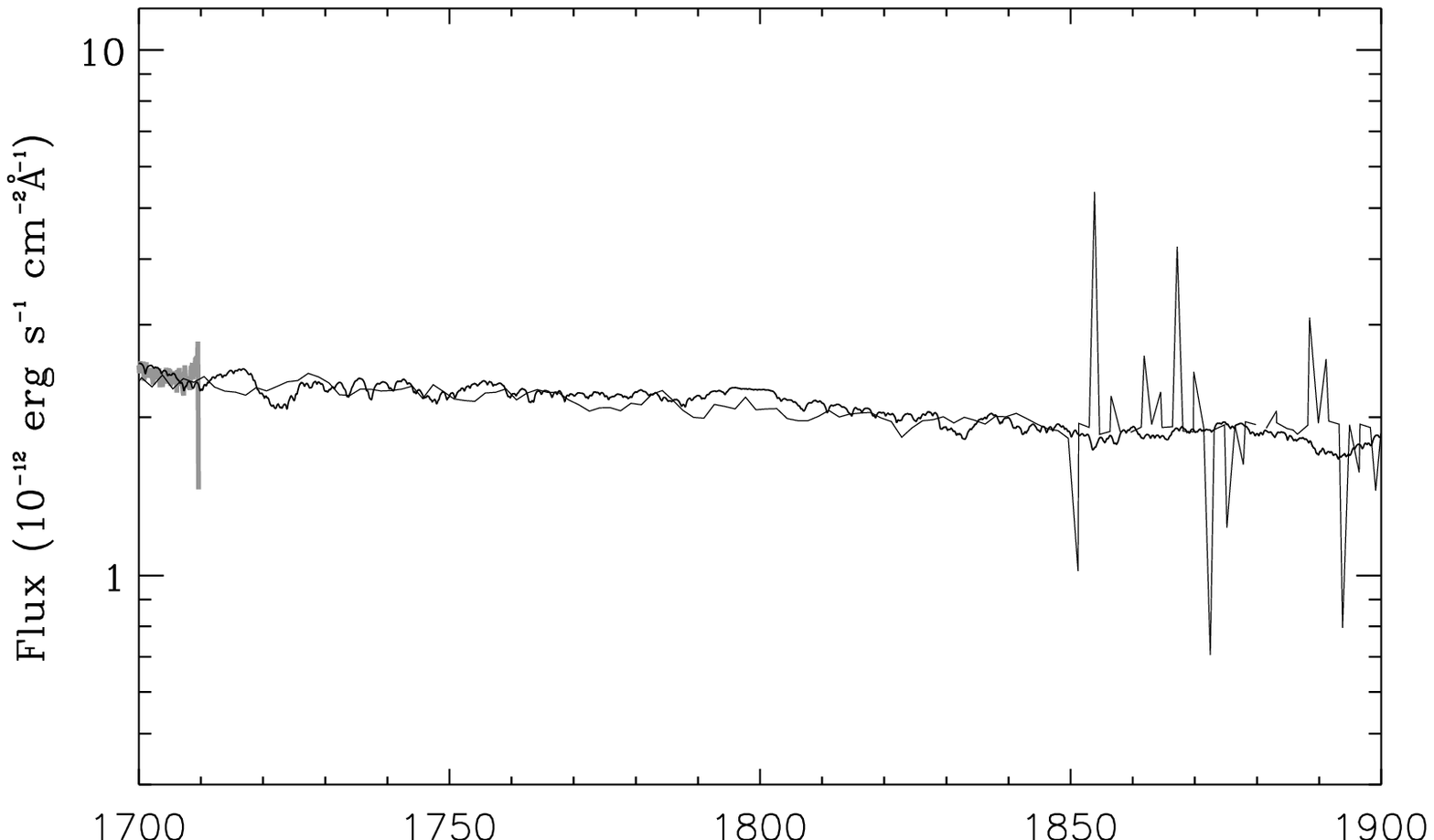}
\epsscale{1.00}
\vspace{6pt}
\figcaption{
Detail of Figure~5 showing fit of the synthetic spectrum (heavy line)
to the long wavelength end of the SWP $IUE$
spectrum (light slowly undulating line) and the short wavelength end 
of the LWP $IUE$ spectrum (light jagged line).
The accurate superposition of the SWP spectrum on the red end of the
$STIS$ spectrum (short, heavy noisy line) is visible at the extreme left of 
the plot.
\label{f9}}
\end{figure}

\begin{figure}[tb]
\epsscale{0.97}
\plotone{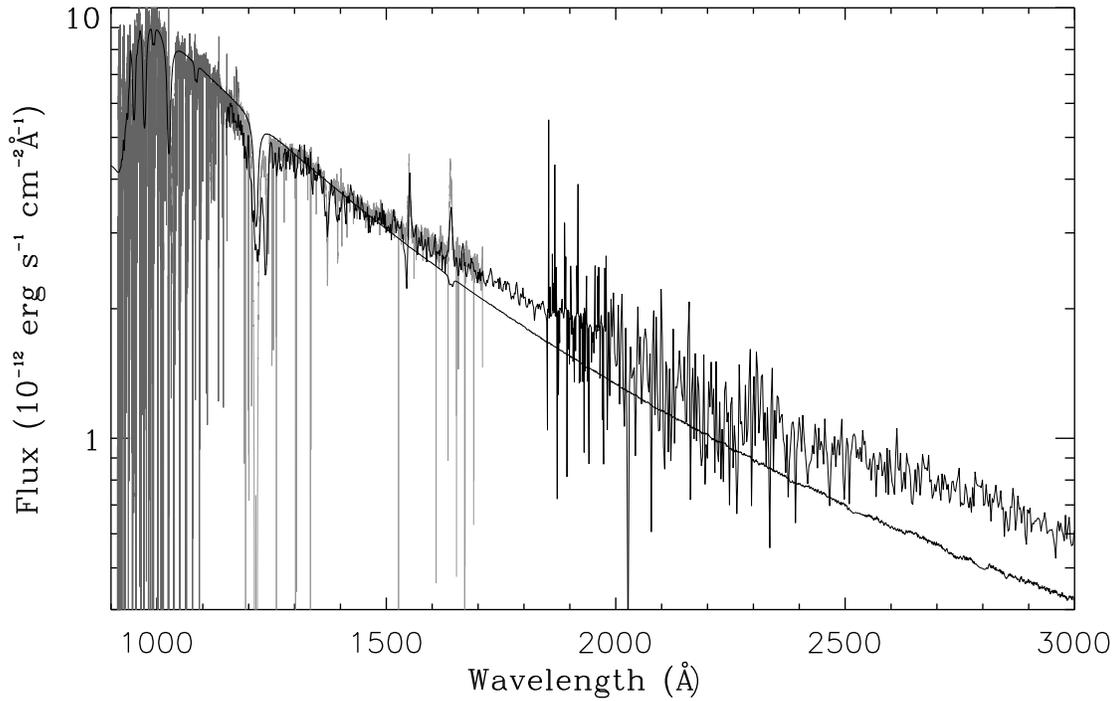}
\epsscale{1.00}
\vspace{6pt}
\figcaption{
Fit of standard model continuum synthetic spectrum to combined observed spectra.
The mass transfer rate is $\dot{M}=3.0{\times}10^{-7}~{M}_{\odot}{\rm yr}^{-1}$
and the WD mass is $0.6M_{\odot}$.
The synthetic spectrum spectral gradient is too large. Compare with Figure~3.
\label{f10}}
\end{figure}

\begin{figure}[tb]
\epsscale{0.97}
\plotone{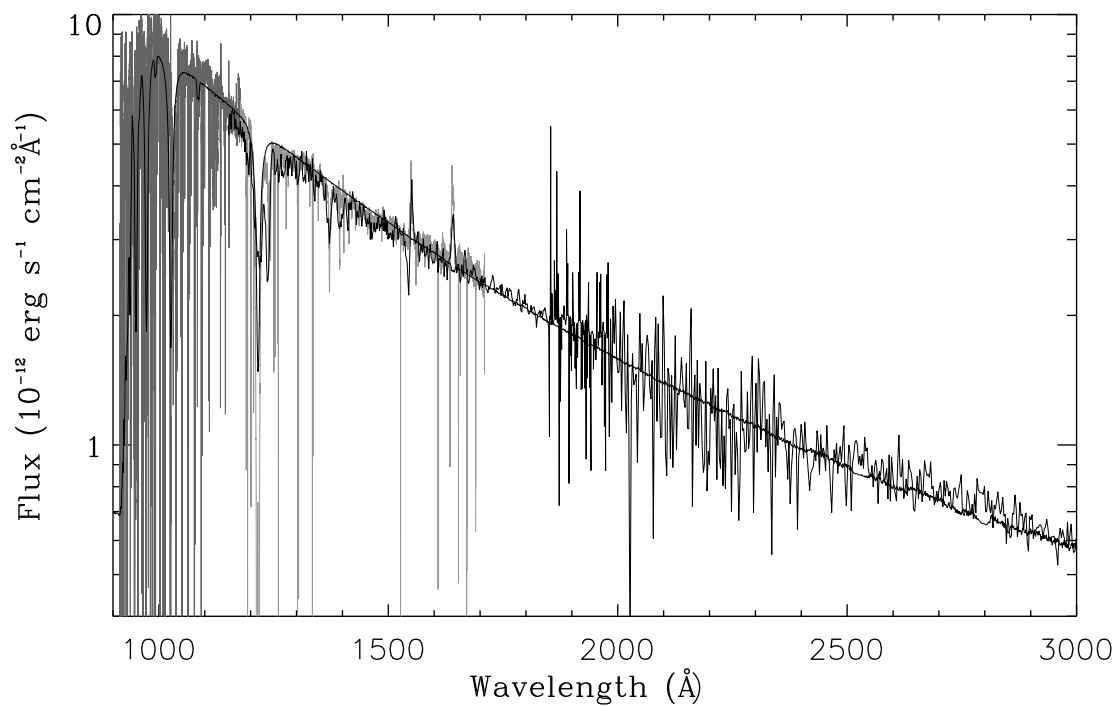}
\epsscale{1.00}
\vspace{6pt}
\figcaption{
Fit of truncated model continuum synthetic spectrum, WD mass of $0.6M_{\odot}$, 
to combined observed spectra.
The mass transfer rate is $\dot{M}=3.0{\times}10^{-7}~{M}_{\odot}{\rm yr}^{-1}$.
The truncation radius is $R_{\rm trunc}=9.91R_{\rm wd}$.
The synthetic spectrum fails to fit the observed spectra 
in the FUV. Compare with Figure~4. 
\label{f11}}
\end{figure}
																							
\begin{figure}[tb]
\epsscale{0.97}
\plotone{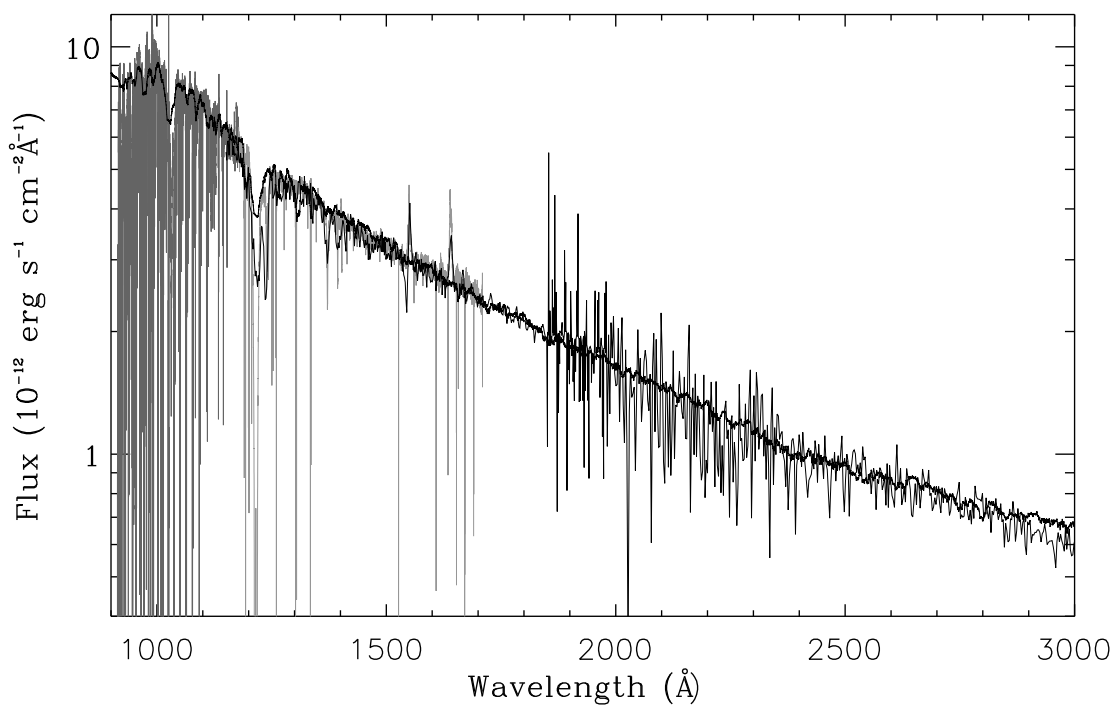}
\epsscale{1.00}
\vspace{6pt}
\figcaption{
Fit of an isothermal model synthetic spectrum (heavy line) to combined observed spectra.
The synthetic spectrum has a resolution of 0.02\AA.
The mass transfer rate is $\dot{M}=3.0{\times}10^{-7}~{M}_{\odot}{\rm yr}^{-1}$
and the WD mass is $0.6M_{\odot}$. The synthetic spectrum has been divided by 
$1.57{\times}10^{42}$. The corresponding distance to QU Car is 406 pc.
See the text for details.
\label{f12}}
\end{figure}

\begin{figure}[tb]
\epsscale{0.97}
\plotone{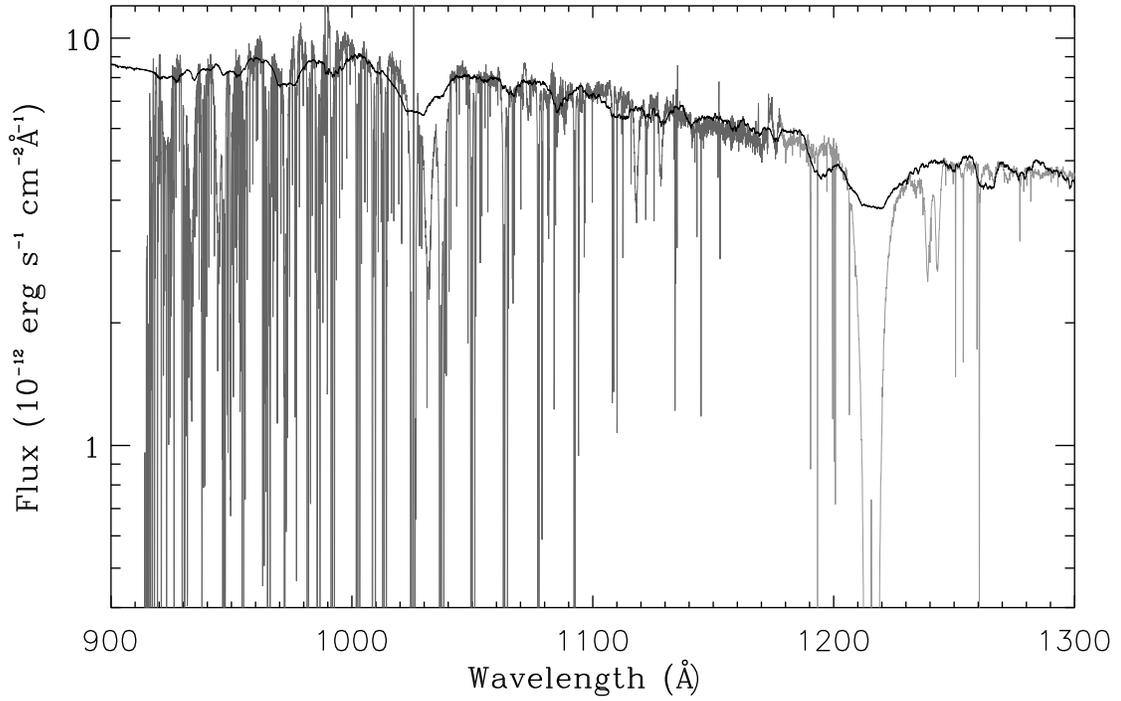}
\epsscale{1.00}
\vspace{6pt}
\figcaption{
Detail of Figure~12 showing comparison with the $FUSE$ spectrum
(grey plot)
and the short wavelength end of the $STIS$ spectrum. Note the
strong ISM absorption lines in the $FUSE$ spectrum.
\label{f13}}
\end{figure}

\begin{figure}[tb]
\epsscale{0.97}
\plotone{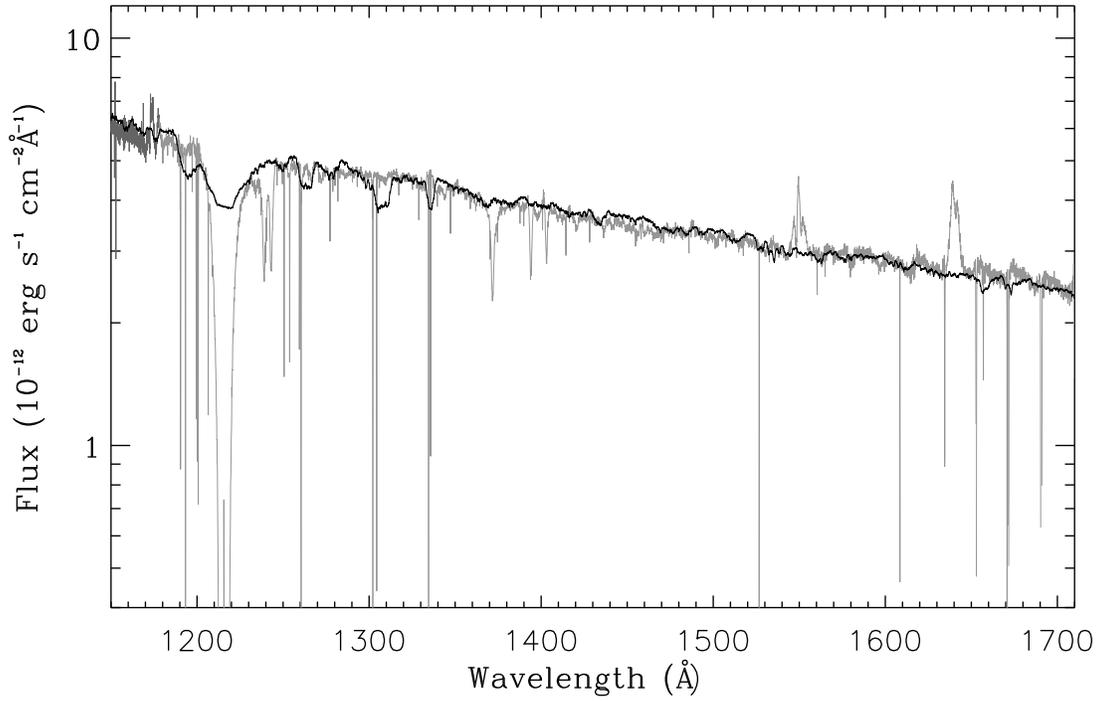}
\epsscale{1.00}
\vspace{6pt}
\figcaption{
Detail of Figure~12 showing comparison with the $STIS$ spectrum.
ISM absorption affects the $STIS$ spectrum shortward of 1500\AA.
The $IUE$ contribution has been suppressed to avoid confusion with the $STIS$
spectrum. 
\label{f14}}
\end{figure}

\begin{figure}[tb]
\epsscale{0.97}
\plotone{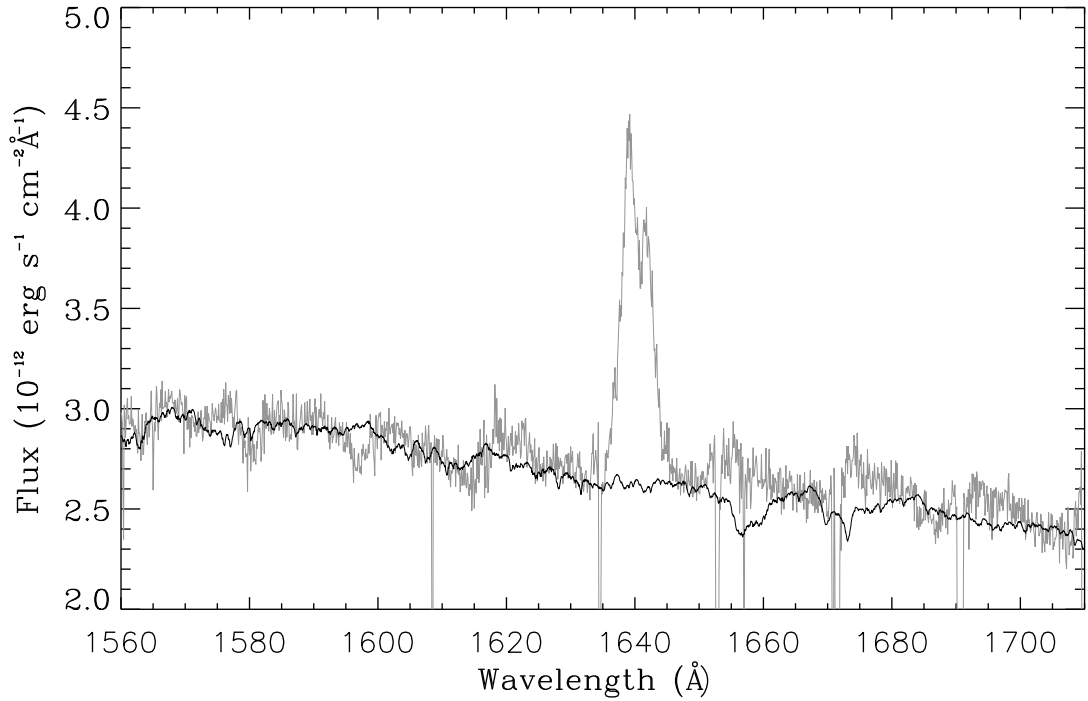}
\epsscale{1.00}
\vspace{6pt}
\figcaption{
Expanded scale detail of Figure~12. Note that the ordinate scale is linear.
The grey line is a segment of the $STIS$ spectrum, resolution 0.1\AA.
The synthetic spectrum resolution is 0.02\AA.
Except for a few details and the He II emission line, the fit is 
quite accurate. Compare with Figure~8. The synthetic spectrum normalizing 
factor is the same as in Figure~12.
\label{f15}}
\end{figure}

\begin{figure}[tb]
\epsscale{0.97}
\plotone{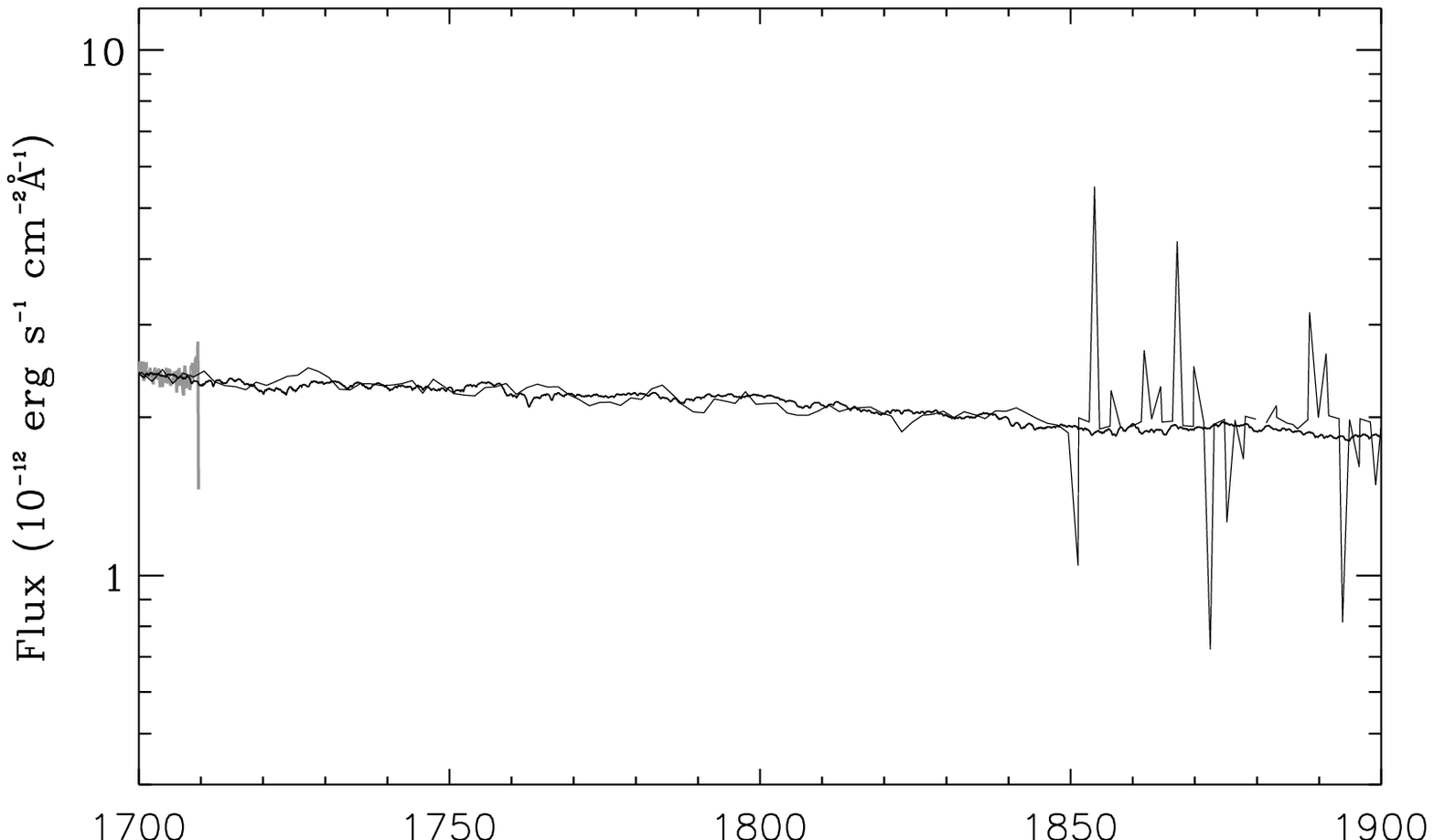}
\epsscale{1.00}
\vspace{6pt}
\figcaption{
Detail of Figure~12 showing the fit of the synthetic spectrum (heavy line)
to the long wavelength end of the SWP $IUE$
spectrum (light slowly undulating line) and the short wavelength end 
of the LWP $IUE$ spectrum (light jagged line).
The accurate superposition of the SWP spectrum on the red end of the
$STIS$ spectrum (short, heavy noisy line) is visible at the extreme left of 
the plot.
\label{f16}}
\end{figure}

\begin{figure}[tb]
\epsscale{0.97}
\plotone{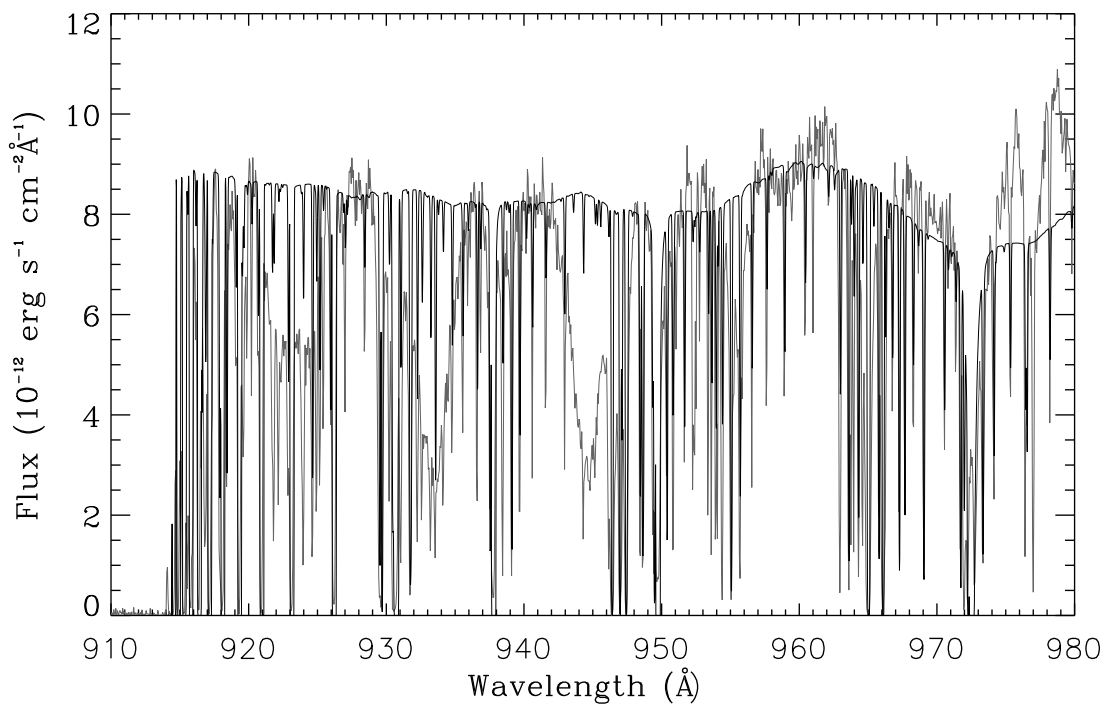}
\epsscale{1.00}
\vspace{6pt}
\figcaption{
ISM-corrected version of Figure~6. 
The spectral resolution of the
synthetic spectrum is 0.02\AA.
Strong absorption lines of N~IV and S~VI (grey underlying plot) remain unmodeled,
possibly from a wind or accretion disk chromosphere. See
Figure~1 for identifications. There now is a good fit to Ly$~\gamma$
and Ly~$\delta$. Note the linear ordinate scale.
See the text
for a discussion.
\label{f17}}
\end{figure}

\begin{figure}[tb]
\epsscale{0.97}
\plotone{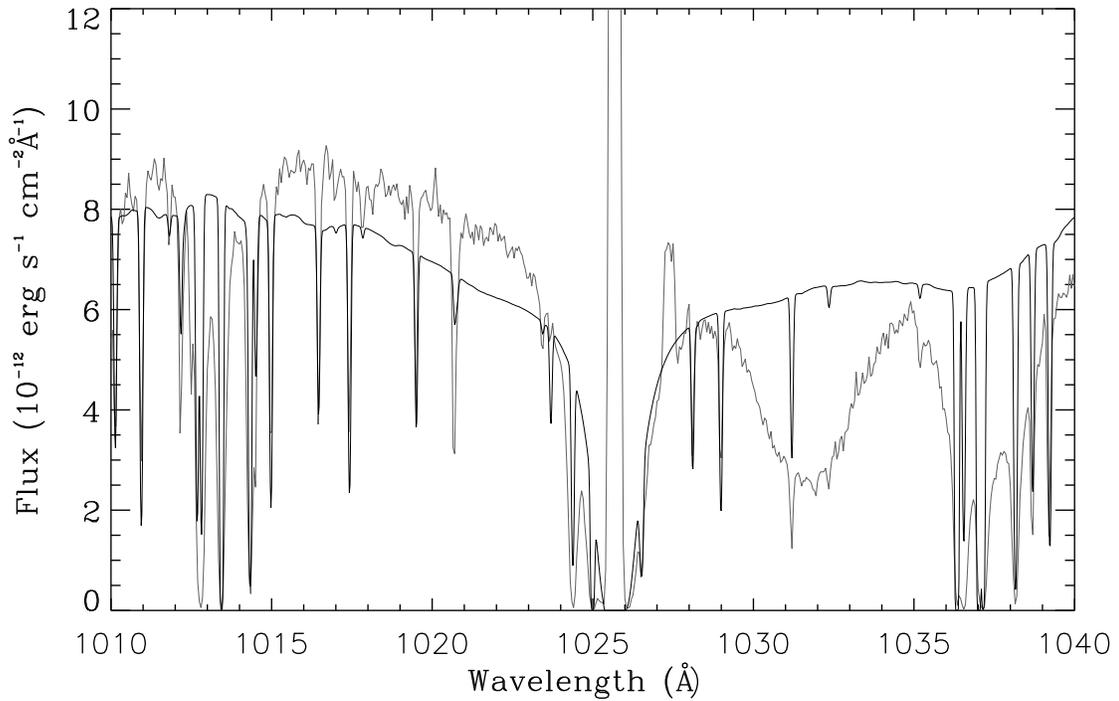}
\epsscale{1.00}
\vspace{6pt}
\figcaption{
Comparison of ISM--corrected synthetic spectrum and $FUSE$ spectrum
near Ly~$\beta$. The ISM correction provides a good fit to a number of 
the narrow ISM lines. Note the linear ordinate scale.
Note the remaining
unrepresented broad O VI lines at 1033\AA~and 1037\AA. These features
probably associate with a wind or chromosphere not included in the model.
\label{f18}}
\end{figure}

\begin{figure}[tb]
\epsscale{0.97}
\plotone{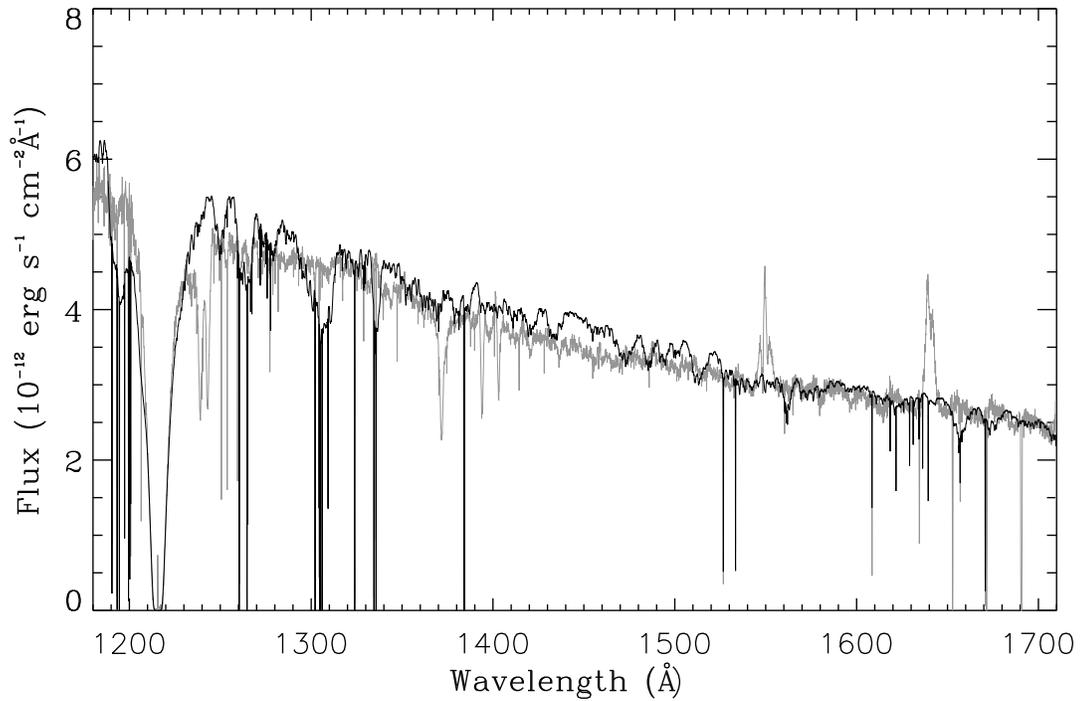}
\epsscale{1.00}
\vspace{6pt}
\figcaption{
Synthetic spectrum fit, including an ISM correction, to $STIS$ spectrum. 
The heavy line is the synthetic spectrum, the grey line
is the observed spectrum.
The $IUE$ contribution has been suppressed to avoid confusion with the $STIS$
spectrum. There now is a good fit to Ly~$\alpha$, in contrast to Figure~7,
but unmodeled features remain. See the text for a discussion. Note the
linear ordinate scale.
\label{f19}}
\end{figure}

\end{document}